\documentclass[preprint, 11pt]{elsarticle}
\usepackage{amssymb}
\usepackage{amsmath}
\usepackage[english]{babel}
\usepackage[utf8x]{inputenc}
\usepackage{graphicx}
\usepackage[colorinlistoftodos]{todonotes}
\usepackage{float}
\usepackage[export]{adjustbox}
\usepackage[top=2 cm, bottom=2cm, left=2.54cm, right=2.54cm]{geometry}
\usepackage{tikz}
\usepackage{array}
\usepackage{url}
\usepackage{indentfirst}
\usepackage{caption}
\usepackage{pbox}
\usepackage{subcaption}
\usepackage{color}

\graphicspath{{./}{./figs/}}

  \def\clap#1{\hbox to 0pt{\hss#1\hss}}

\usepackage{bm}
\providecommand{\mat}[1]{\bm{#1}}%
\renewcommand{\vec}[1]{\mathbf{#1}}

\providecommand{\mC}{\ensuremath{\mat{C}}}

\providecommand{\mW}{\ensuremath{\mat{W}}}

\providecommand{\mLambda}{\ensuremath{\mat{\Lambda}}}

\providecommand{\vw}{\ensuremath{\vec{w}}}
\providecommand{\vx}{\ensuremath{\vec{x}}}

\newcommand{\bmat}[1]{\begin{bmatrix}#1\end{bmatrix}}

\newcommand{\diag}[1]{\operatorname{diag}(#1)}

\usetikzlibrary{shapes,shadows,arrows}
\tikzstyle{decision} = [rectangle, draw, fill=blue!20, 
    text width=10em, text badly centered]
\tikzstyle{block} = [rectangle, draw, fill=magenta!20, 
    text width=10em, text centered, rounded corners, minimum height=4em]
\tikzstyle{block2} = [rectangle, draw, fill=blue!20, 
    text width=10em, text centered, rounded corners, minimum height=4em]
\tikzstyle{line} = [draw, -latex]
\tikzstyle{cloud} = [draw, ellipse,fill=red!20, node distance=1cm,
    minimum height=2em]
\tikzstyle{empty} = [node distance=4cm]

\journal{Applied Mathematics and Computation}

\begin{document}

\begin{frontmatter}

\title{A Modified SEIR Model for the Spread of Ebola in Western Africa and Metrics for Resource Allocation}
%Paul Constantine,
\author{Paul Diaz}
\author{Paul Constantine}
\author{Kelsey Kalmbach}
\author{Eric Jones}
\author{Stephen Pankavich\fnref{Grants}}

\fntext[Grants]{The authors were supported by the joint Mathematical Association
    of America (MAA) and Society for Industrial and Applied Mathematics (SIAM) PIC Math program funded by the National Science Foundation (NSF grant DMS-1345499).  Additionally, the last author was supported by the National Science Foundation under award DMS-1211667.}
    
\address{Department of Applied Mathematics and Statistics\\
Colorado School of Mines\\
1500 Illinois St.\\
Golden, CO 80401}
\ead{pankavic@mines.edu}

% Journals - JMB, PLoS Current Outbreaks,
% International Journal of Epidemiology, American Journal of Epidemiology,
% Journal of Clinical Epidemiology, Journal of Epidemiology and Community Health

%\begin{document}
%\maketitle 

\begin{abstract}
A modified, deterministic SEIR model is developed for the $2014$ Ebola epidemic occurring in
the West African nations of Guinea, Liberia, and Sierra Leone.  The model describes the dynamical interaction of susceptible and infected populations, while accounting for the effects of hospitalization and the spread of disease through interactions with deceased, but infectious, individuals.
Using data from the World Health Organization (WHO), parameters within the model are fit to
recent estimates of infected and deceased cases from each nation.  The model
is then analyzed using these parameter values.  Finally, several metrics are
proposed to determine which of these nations is in greatest need of additional 
resources to combat the spread of infection. These include local and global sensitivity metrics of both the infected population and the basic reproduction number with respect to rates of hospitalization and proper burial.
\end{abstract}


\end{frontmatter}

%%%***Possible Reviewers - Sara Del Valle (LANL), Jan Medlock (Oregon State)

\section{Introduction \label{background}} 
%(see Appendix \ref{fig:days_since} Figure 6). 
%
\noindent The most recent Ebola outbreak began in December 2013 and resulted in a devastating loss of life in Guinea, Liberia, and Sierra Leone. This outbreak has been the deadliest in the history of the disease claiming more than $10,000$ lives to date~\cite{cdc, whositrep}. The severity of the epidemic has prompted an international response to halt further spread of the disease. In particular, efforts arising within the United States and China have provided additional relief to the region, but within 2014 this had not been enough to end the epidemic. 

Some basic facts about Ebola disease pathogenesis are well known. Those exposed to the virus experience an 8-to-10 day average incubation period during which they remain noninfectious \cite{cdc}, though the time between exposure and the onset of symptoms ranges from $2$ to $21$ days~\cite{dixon2014ebola}. Once a patient becomes symptomatic, the virus may be transferred to others through direct contact with bodily fluids \cite{DuToit}, such as blood and vomit. One important epidemiological feature of Ebola is that those who are killed by the disease can still transmit the virus to susceptible individuals.
Unlike most pathogens, which cannot survive long on a deceased individual, Ebola does remain infectious after a person succumbs to the disease~\cite{dixon2014ebola}.
In fact, the deceased are even more contagious than living Ebola patients as the virus can force a victim's body to release infectious fluids including blood, vomit, and fecal matter - especially in later stages of the disease ~\cite{CDC1}. This fluid release is among the most visually harrowing symptoms of Ebola, present in some late-stage patients~\cite{dixon2014ebola}.

The spread of infection from deceased to susceptible individuals is a significant problem in Western Africa where local burial rituals often require washing, touching, or kissing the body of the deceased. 
Among the traditional practices the WHO cautions against with Ebola victims are family-led
body preparation and religious rituals that require direct contact with the corpse. Muslim tradition, for instance, requires that family members of the same gender wash the body themselves before burial.
Moreover, the lack of adequate health care within Guinea, Liberia, and Sierra
Leone has perpetuated the disease, as improper and unsafe burials place further
individuals at risk. Widespread reports from Liberia in late 2014 described Ebola victims laying on the streets for days, drastically increasing the risk of infection~\cite{WHO1}. The WHO estimates that contact with deceased individuals has caused at least $20\%$ of all infections~\cite{WHO2}. 
Hence, the bodies of deceased Ebola victims that have not been properly disposed and
burial ceremonies, in which mourners have direct contact with the body of the deceased, can play a large role in the transmission of the virus, and these effects cannot be neglected within an informative model.

Another important feature is the effect of hospitalization.
Due to their frequent interaction with patients either under investigation or confirmed to have contracted the Ebola virus, healthcare personnel are particularly susceptible to the disease \cite{WHO3}.
In particular, they can be exposed to Ebola by coming in contact with a patient’s body fluids, contaminated medical supplies and equipment, or contaminated environmental surfaces.  For these reasons transmission between infectious, hospitalized individuals and medical workers are non-negligible within a descriptive epidemic model. Additionally, hospitalization and subsequent treatment does improve an infected patient's chances of recovery, though the difference in case fatality rates between hospitalized patients and non-hospitalized cases is relatively small \cite{NEJM}, and an increase in the use of medical facilities within a population can reduce the damage caused by the epidemic. 

While previous studies have focused predominantly on the effects of contact tracing \cite{Rivers, webb2014model}, it has recently been determined \cite{Science} that other aspects of disease pathogenesis, including hospitalization, case isolation, and further introduction of sanitary funeral processes, must be better addressed in order to fully mitigate further spread of the disease.
Hence, the ultimate goal of the current study is to construct a model for the most recent Ebola epidemic in Western African and identify useful metrics to determine which of the affected countries would benefit most from the allocation of additional resources, including new treatment facilities or greater manpower to ensure proper burials.  To this end, a deterministic model for the epidemic is constructed in the next section by accounting for the specific characteristics of the disease, including the incubation period, increased risk of infection from deceased individuals, and the effects of hospitalization. 
Realistic parameter values are determined by fitting to given data for each of the nations and a separate analysis of the respective reproduction numbers is conducted. In Section $3$ we discuss local metrics other than direct computation of the basic reproduction number ($R_0$) which may allow one to determine the optimal sites for resource allocation and inhibit the spread of the disease. 
%In particular, due to the constraint of limited resources, a fundamental interest is to determine which of these nations would most benefit from the placement of a new medical treatment facility to increase the proportion of infected individuals who are hospitalized.  
These metrics include fitted parameter values, the local sensitivity of the infectious population, and the local minimization of $R_0$ with respect to parameter variation.  Within Section $4$ we propose and explore a new method to compute the global sensitivity of $R_0$ with respect to alterations in the most crucial parameter values.
Ultimately, we find that of these three troubled nations, it is Liberia that would benefit most from the introduction of additional medical resources. In general, these tools - especially the proposed activity scores of Section $4$ - can be used to analyze future outbreaks in addition to the most recent crisis. Finally, the relevant details of computations are contained within an appendix.

\section{A Mathematical Model for Ebola}
\label{sec:two}
\noindent Our first objective is to develop a mathematical model for the spread of Ebola in Western Africa by accounting for the specific characteristics of the disease. 
SEIR models are often implemented when studying the spread of infectious diseases that possess significant incubation periods \cite{ABG, Legrand, Rivers, SYK, webb2014model}.
To describe the spread of Ebola, a traditional SEIR model is augmented with additional compartments based on the following assumptions.
First, as the number of susceptibles is large, we neglect stochastic effects and formulate a deterministic model.
Additionally, though the current epidemic has lasted for over a year, we assume that the outbreak is not sustained by the introduction of new susceptible individuals.  Hence, national populations are near equilibrium and both births and deaths are neglected amongst the total population.
We further assume that infected individuals can move to three different removed compartments: removed and infectious (i.e. not properly buried), removed and properly buried, and removed and recovered.
This distinction is introduced to allow for the scenario in which individuals who have died from the disease, but have not been properly buried, may continue to infect susceptible individuals whom they contact.

We assume that those who recover from the disease are no longer susceptible, as survivors of Ebola are thought to be immune to the strain of the virus that infected them \cite{CDC1}.
Once hospitalized, infected individuals can still spread the disease to members of the susceptible population. However, we assume that patients who die within a hospital receive an immediate proper burial, and thus cannot infect others once deceased.
Finally, we assume that hospitalized individuals have a greater chance of survival than infected, non-hospitalized individuals~\cite{dixon2014ebola}. This is consistent with estimates of the general Ebola fatality rate around $70\%$ and the hospitalized fatality rate near $64\%$~\cite{NEJM}.

\subsection{A Modified SEIR Model}

\noindent With the aforementioned assumptions identified, we formulate a modified SEIR model to account for the dynamics of the disease within a population. The model consists of the following seven states, each a function of time $t$: (i) the susceptible proportion of the population $S(t)$, (ii) the exposed proportion (i.e., infected but asymptomatic) $E(t)$, (iii) the infected and symptomatic proportion $I(t)$, (iv) the infected and hospitalized proportion $H(t)$, (v) the removed but infectious proportion (i.e., those who died from the disease but have not been sanitarily buried) $R_I(t)$, (vi) the removed and buried proportion $R_B(t)$, and (vii) the removed and recovered proportion $R_R(t)$.

There are several ways in which the populations may be altered. First, susceptibles transfer to the exposed population after coming into contact with either an infected individual (including those who are hospitalized) or a body which is not yet buried. After the viral incubation period, all exposed individuals move to the infected population. The infected move to one of three populations: hospitalized, removed and infectious, or removed and recovered. Note that infected individuals cannot move immediately to a removed and buried state since we assume that some time is needed to bury an individual not receiving medical care at death. However, a proportion of removed and infectious individuals transfer to removed and buried. Hospitalized individuals move to either removed and buried or removed and recovered. Finally, all transmission terms are assumed to be of standard incidence form. Figure \ref{tikzCM} summarizes these transition pathways between populations, while Table \ref{tab:paramdescrip} summarizes model parameters---all of which are nonnegative.

The coupled system of differential equations for $S(t)$, $E(t)$, $I(t)$, $H(t)$, and $R_I(t)$ is given by
\begin{equation} \left.
\begin{aligned} 
\frac{dS}{dt} &= - \beta_1 S I -\beta_2 S R_I -\beta_3SH\\
\frac{dE}{dt} &=  \beta_1 S I +\beta_2 S R_I +\beta_3SH- \delta E \\
\frac{dI}{dt} &=  \delta E - \gamma_1 I-\psi I\\
\frac{dH}{dt} &= \psi I - \gamma_2 H \\ 
\frac{dR_I}{dt} &= \rho_1\gamma_1 I - \omega R_I.
\end{aligned} \label{std_seir}
\right \}
\end{equation}
The $R_B$ and $R_R$ proportions decouple from the system above as their values are determined once the others are known. Their respective time evolution is given by
\begin{equation} \left.
\begin{aligned} 
\frac{dR_B}{dt} &= \omega R_I+\rho_2\gamma_2 H \\ 
\frac{dR_R}{dt} &= (1-\rho_1)\gamma_1 I+(1-\rho_2)\gamma_2 H.
\end{aligned} \label{std_R}
\right \}
\end{equation}
With $R_B$ and $R_R$ accounted for, the model is conservative, and compartments have been rescaled so that each represents the proportion of a specific population with respect to the total population.  For example, denoting the total population constant by $N = S+E+I+H+R_I+R_B+ R_R$ so that $\frac{dN}{dt} = 0$, we can express the first unknown function as $S(t) = \frac{s(t)}{N}$ where $s(t)$ represents the total number of susceptible individuals within the nation of interest.
Our model is similar to a stochastic SEIR model derived in \cite{Legrand} to explain previous outbreaks in Uganda, Gabon, Sudan, and the Democratic Republic of the Congo.  However, we employ a deterministic model herein and specifically assume that, due to current public health intervention, any individuals who die while hospitalized are properly buried.
Additionally, the removed population within our model is separated into deceased (and noninfectious) and recovered individuals in order to provide more reasonable estimates of the impact of the disease.

\subsection{Parameter Fitting} \label{sec:meth}
\noindent We begin our analysis of the model by identifying values of model parameters that generate predictions matching available data for Guinea, Liberia, and Sierra Leone. We obtained time-series data for each nation from WHO Situation reports~\cite{whositrep} and the Network Dynamics and Simulation Science Laboratory at Virginia Tech~\cite{dataset}. The data sets contain cumulative values of infections and deaths from each nation. However, the data are incomplete and/or irregularly reported; hence, we remove outliers and time periods without sufficient reporting. The remaining number of data points are 36, 90, and 61 for Liberia, Guinea, and Sierra Leone, respectively.

Since the data represent cumulative quantities, but the model describes instantaneous proportions of active infections, and differing compartments are employed in our model for deaths, a direct fit is not immediately possible.  Instead, to generate cumulative quantities, the time-integrated infected population was fit to the cumulative infected data. 
Also, our model includes three deceased states, while the data do not differentiate among differing death compartments. Thus, we assume reported deaths in the data correspond to the properly buried population $R_B$, and not the deceased but infectious population $R_I$, as the latter are likely unknown to data collectors. 

To fit the model parameters, an unconstrained nonlinear optimization was performed using MATLAB's \texttt{fminsearch} function, which utilizes a Nelder-Mead direct search method. Within this solver nonlinear parameter constraints were enforced by using a barrier function to ensure positive, realistic parameter values.
%~\cite{lagarias1998convergence}. 
The objective function, $D(\mathbf{p})$, where $\mathbf{p}$ represents the vector of parameters listed in Table \ref{tab:paramdescrip}, is defined as
\begin{equation}
\label{eq:err}
D(\mathbf{p}) \;:=\; \sum_{t \in \mathcal{T}} [R_{data}(t) - N\cdot R_B(t; \mathbf{p})]^2 + \left[C_{data}(t) -  N \cdot \delta\int_0^t E(s;\mathbf{p})\,ds \right]^2 
\end{equation}
where $\mathcal{T}$ is the discrete set of times at which the data is available, $N$ is the total population for a given nation, $R_{data}(t)$ is the recorded number of cumulative deceased individuals, and $C_{data}(t)$ is the recorded number of cumulative infections. 
The initial conditions are the proportions as of March 22, 2014.  For example, the total estimated population of Liberia is $N=4.29$ million, while the number of reported infections on this date was $9$, and hence $I(0) = 9 / (4.29\times 10^6) = 2.1 \times 10^{-6}$.

Table \ref{tab:Param} displays the fitted parameter values for each independent country. 
%---fit with least squares using initial guesses from the sources in the last column. 
Estimates of the incubation period range between $8$ and $10$ days, so the daily probability of transition from the exposed state to the infected state was assumed to be $\delta = \frac{1}{9}$ for simplicity. 
Figure \ref{fig:paramgraphs} displays model trajectories with fitted parameters compared to the available data for Guinea, Liberia, and Sierra Leone.

With parameter values established, the basic reproduction number (i.e., the average number of secondary infections generated by a representative primary case within an entirely susceptible population) can be computed for each nation.  Using the next generation matrix method (see Appendix), this quantity is found to be
\begin{equation}
\label{R0}
R_0 = \frac{\beta_1 + \frac{\beta_2 \rho_1 \gamma_1}{\omega} + \frac{\beta_3}{\gamma_2}\psi}{\gamma_1 + \psi}.
\end{equation}
We note that this quantity is independent of the parameters $\delta$ and $\rho_2$ since neither the length of the incubation period nor the death rate of hospitalized patients affects the generation of secondary infected cases.
Utilizing fitted parameter values the basic reproduction number for each nation was computed, as expressed within Table \ref{tab:R0}.  Liberia appears to be experiencing the greatest detrimental effects of the epidemic, as evidenced by this metric, but the corresponding value for Sierra Leone is nearly as large, while Guinea does not differ too dramatically from the others.
Without further intervention and new allocation of resources to fight the disease, our model predicts that the epidemic would continue to spread as of December $2014$, generating a large number of new cases each day - $23$ in Guinea, $119$ in Sierra Leone, and $736$ in Liberia.

\section{Local Metrics for Resource Allocation}
\noindent Given the limited amount of resources available to these West African nations and the diminishing availability of external aid from charitable organizations around the world, a natural question to investigate is which of the nations currently experiencing an epidemic would benefit most from additional hospitalization and treatment resources.
Thus, another major goal of this work is to study different metrics by which such decisions can be accurately established.  Certainly, one way to answer this question is to compare the computed basic reproduction numbers for each region and allocate new resources to the one whose $R_0$ value is largest.  However, this single value may not always capture a necessary level of detail within infection dynamics and doesn't account for parameter variation due to intervention.  Hence, in this section we investigate both changes to $R_0$ and a few other possible metrics to determine the placement and allocation of new resources.

\subsection{Fitted Parameter Values}\label{subsec:fits}

\noindent Computing some of the specific mean times and probabilities (see Table \ref{tab:Spef}) can provide a preliminary understanding of the epidemic. These quantities were derived directly from estimates of fatality rates provided by \cite{NEJM} and the parameter values (Table \ref{tab:Param}) fit to the data.  For instance, the average time from patient hospitalization to death was computed using the parameters $\rho_2$ and $\gamma_2$.  Namely, if $t_{HD}$ represents this average time, and the fatality rate of all hospitalized cases in Liberia is $67\%$, as reported in \cite{NEJM}, then the term $\rho_2 \gamma_2$ within \eqref{std_seir} can be represented as the product of the probability that a hospitalized patient dies and the average rate at which a patient exits the hospitalized population due to death, or 
$$\rho_2 \gamma_2 = 0.67\cdot \frac{1}{t_{HD}}.$$
Using the fitted values of $\rho_2$ and $\gamma_2$ in Liberia, one can uniquely determine $t_{HD}$.
In the same way, the fit parameters $\rho_1, \gamma_1, \psi$, and $\omega$ were used to determine the remaining stated mean times for Liberia and Sierra Leone.  
These rates were not computed for Guinea as the behavior of the epidemic in this nation differs dramatically from the others, as noted in \cite{Rivers}.
Because of this difference and the comparatively small value of $R_0$, we will disregard the spread of the disease within Guinea for the remainder of the paper.

Notice that Sierra Leone was determined to possess the longest average time from hospitalization to recovery and a larger percentage of hospitalized cases, while Liberia experiences the longer wait for proper burials on average.
The two nations experience similar wait times for hospitalization. 
We further note that neither possesses uniformly larger infection rates, as transmission from the infected and hospitalized populations is largest in Liberia, but transmissions from deceased individuals is greatest in Sierra Leone.
In addition, the number of infectious individuals and the rate of growth of the deceased population is largest within Liberia (Fig. \ref{fig:paramgraphs}).
Other quantities, such as the proportion of the total population stricken with the disease, could also be considered.
In general, parameter values alone, even when determined directly from the infected and deceased data sets, likely cannot determine the country most in need of additional resources. 
%Based on these comparisons it is unclear which nation is in the greatest need of additional resources.
		
\subsection{Greatest Reduction in $R_0$}	

\noindent Since an analysis of the basic reproduction number is a traditional metric to characterize the spread of disease, it is a natural quantity to consider reducing through intervention.  In Section \ref{sec:two}, values of $R_0$ were computed for each of the three nations.  As a function of parameters within the model, the basic reproduction number is expressed by \eqref{R0}.  

Until new pharmaceutical and enhanced medical treatments are developed, it appears that the parameters $\rho_1, \rho_2, \gamma_1$, and $\gamma_2$ cannot be greatly altered by the allocation of additional resources.  Similar statements hold true for the infection parameters ($\beta_1, \beta_1, \beta_3$) and incubation period ($\delta$), which are determined inherently by population interactions and the disease itself, though infection rates can be altered by changes in human behavior.  The parameters over which one has the greatest immediate control and are most strongly altered by the addition of human resources are the hospitalization parameter $\psi$ and rate of proper burial $\omega$.  Hence, we may fix the other parameters in the system and view $R_0$ as a function of $\psi$ and $\omega$.
Then, the local rate of change of $R_0$ with respect to these parameters should provide insight into the efficacy of an increase in either resource.
Using \eqref{R0}, we compute
$$\frac{\partial R_0}{\partial \psi} = \frac{- \beta_1 - \beta_2\frac{\rho_1 \gamma_1}{\omega} + \beta_3\frac{\gamma_1}{\gamma_2}}{(\gamma_1 + \psi)^2}$$
and note that this quantity is negative if the numerator is negative. 
With parameter values from Table \ref{tab:Param}, the derivative is negative for each of the three nations, as one would expect a boost in the hospitalization rate $\psi$ to lower the reproduction number.  
Similarly, we find
$$\frac{\partial R_0}{\partial \omega} = -\frac{\beta_2\rho_1 \gamma_1}{(\gamma_1 + \psi) \omega^2},$$
and because all parameters are positive, this quantity must be negative as well.
At this point, we may use these partial derivatives to gauge the local sensitivity of $R_0$, but a more descriptive quantity is the normalized forward sensitivity index \cite{CHC}, namely the ratio of the relative change in $R_0$ to the relative change in the parameter of interest, defined by 
$$\Upsilon^{R_0}_p : = \frac{\partial R_0}{\partial p} \cdot \frac{p}{R_0}$$
where $p = \{\psi, \omega\}$.
The quantities of interest for a nation are then the values of these normalized sensitivity indices, and we are particularly interested in how such values differ from their baselines, i.e. $\Upsilon^{R_0}_p$ evaluated at the current values of $\psi$ or $\omega$, which we denote by $\psi_0$ and $\omega_0$ respectively. Computed values of these indices are displayed in Table \ref{tab:R0prime_2param}.
Additionally, contour plots of $R_0(\psi, \omega)$ for the two nations are displayed in Figure \ref{fig:contours}.
They clearly demonstrate that  the influence of $\psi$ is much greater than that of $\omega$.
Hence, an increase in the hospitalization of infected individuals would have a significantly greater impact on the reproduction number than devoting more resources to ensuring quick and proper burials of the deceased. %Correspondingly, the influence of either variable within Guinea is small compared to the effects in the other two nations.
Therefore, we reduce the problem to viewing $R_0$ as a function of $\psi$ alone and investigate the relative impact that an increase in $\psi$ may have on $R_0$.  
Hence, we fix $\omega = \omega_0$ and compute $R_0'(\psi_0)$ and for each country, the values of which are represented within Table \ref{tab:R0prime}.
Of course, the values of $\Upsilon^{R_0}_\psi$ in Table \ref{tab:R0prime_2param} arise from those of $R_0'(\psi_0)$ in Table \ref{tab:R0prime}.
Since $R_0^\prime(\psi) < 0$ for all $\psi \geq 0$, the minimal value of $R_0$ is obtained by
$$\lim_{\psi \to \infty} R_0(\psi) = \frac{\beta_3}{\gamma_2}$$
while the maximal value occurs when no medical intervention is present and is
$$R_0(0) = \frac{\beta_1}{\gamma_1} + \frac{\beta_2 \rho_1}{\omega}.$$
These respective values are also provided within Table \ref{tab:R0prime}.
%While the specific values of $R_0$ in Table \ref{tab:R0prime_2param} suggest that the three nations are similarly in need of aid, the information described by $R_0'(\psi)$ suggests that additional medical resources to boost the hospitalization parameter, and hence reduce the average time to hospitalization, will not significantly impact Guinea.  Instead, the greatest impact can be made by allocating new facilities within Liberia and Sierra Leone.
The information described by $R_0(\psi)$ and $\Upsilon^{R_0}_\psi$ suggests that additional medical resources to boost the hospitalization parameter, and hence reduce the average time to hospitalization, will significantly impact both Liberia and Sierra Leone, with a slightly greater benefit to the former nation in comparison to the latter.

\subsection{Local Sensitivity of the Infected Population}
\noindent Another possible metric is the change in the infected population as a function of the hospitalization parameter $\psi$. As mentioned within the previous section, this parameter appears to be one of the most controllable factors of the disease's progression, and as we will show in the next section, it is the most influential in terms of the value of $R_0$. Hence, to conduct a local sensitivity analysis of the infected population as a function of $\psi$ only, we fix all model parameters to their fitted value except for $\psi$. In particular, since this function $I(t; \psi)$ is also time-dependent we must rely on simulations of the system to approximate changes in this output over a period of time. This method was implemented with changes in $\psi$ made at  two differing points in time for Liberia and Sierra Leone. In both implementations the parameter $\psi$ is varied between one half and twice its fitted value.

The first implementation focuses on changes in $I(t; \psi)$ arising from varying $\psi$ at time $t=0$, the beginning of the outbreak. Figure \ref{fig:sensitivity} depicts the results of this analysis. In both countries, we see that increased values of $\psi$ correspond to decreases in $I$. However, increasing $\psi$ in Liberia produces a greater decrease in the infected population compared to that of Sierra Leone. This is evident due to the scales of the graphs in Figure \ref{fig:sensitivity}. The second implementation focuses on changes in $I(t;\psi)$ arising from altering the hospitalization parameter at the final recorded time of the data set, thereby simulating the implementation of new treatment strategies at the most recent time. The initial conditions used for future forecasting are given by the last day of the data set. For each value of $\psi$ we forecast the model fifty days into the future.  Figure \ref{fig:future_psi_sensitivity} summarizes the forecasting sensitivity analysis. Again, within both countries we see that increasing $\psi$ reduces $I(t,\psi)$, but the reduction in Liberia is much greater than in Sierra Leone. 
In each of these cases, the local sensitivity of cumulative, rather than instantaneous, infections was also investigated and yielded similar results.

Hence, from this metric we conclude that variations in the hospitalization rate have a significant impact on the rate of infection within both Sierra Leone and Liberia. However, the largest impact is in the country of Liberia. A $30$ percent increase in $\psi$ leads to a nearly $1000$ fewer infectious individuals in Liberia by the end of the forecasting simulation, while the same increase in Sierra Leone would result in a couple hundred fewer infectious individuals.

\section{Global Metrics for Resource Allocation}
\noindent Though the local sensitivity analyses of the previous section were informative, none effectively captures the effects of large variations in model parameters on the quantity of interest.  This occurs because evaluation of the partial derivatives of the reproduction number with respect to parameters must be performed at a specific point $\mathbf{x}$ in the input parameter space. Within the context of the Ebola model, input parameters can be represented by $\mathbf{x} = (\beta_1,\beta_2,\beta_3,\rho_1,\gamma_1,\gamma_2,\omega,\psi) \in \mathbb{R}^8$, while the input parameter space $\Omega\subset \mathbb{R}^8$ is defined by the ranges of each parameter in $\vx$ as summarized in Table \ref{tab:intervals}. Of course, the previous methods measured local sensitivity because model input parameters were perturbed about a specific point $\mathbf{x}\in \Omega$ in order to understand their effect on variations in the model output, e.g. $\frac{\partial R_0}{\partial \psi} (\vx)$. Contrastingly, \emph{global} sensitivity metrics account for parameter variation across the entire input space $\Omega$. Much like local measures of sensitivity, investigating the global effects of input parameters can allow us to establish a relative notion of importance among the model inputs. One class of global sensitivity methods, known as \emph{active subspace} methods, are particularly useful for models with a scalar quantity of interest  \cite{constantine2015active,diaz2015global}.
As  the scalar reproduction number $R_0$ is a particular quantity of interest in epidemiological models and this metric is a strong indicator of the behavior of the disease, we will utilize active subspaces to perform a global sensitivity analysis of $R_0(\vx)$.

To simplify our calculations we first shift and scale the parameter space of dimension eight to the hypercube $ \Omega' = [-1,1]^8$ and equip it with a uniform probability density, $\rho(\vx) = 2^{-8}$. The choice of a uniform density limits our assumptions and details the lack of \emph{a priori} knowledge of the most likely parameter values, as we are assigning equal likelihood to all values within the given range. 
Of course, as additional statistical research is conducted concerning parameters in the model (e.g., transmission, fatality, and recovery rates), a more comprehensive understanding of the distributions of these parameters will be obtained, and $\rho(\vx)$ can be easily altered to better describe our current knowledge of disease dynamics.

Given the normalized parameter space and the associated density, active subspaces of the parameter space are defined in terms of the eigenpairs of the $8\times8$ positive, semi-definite matrix
\begin{equation}\label{eq:C} \mC := \int_{\Omega'} \nabla R_0(\vx) \nabla R_0^T(\vx) \rho(\vx) d\vx \;=\;
\mW\Lambda\mW^T, 
\end{equation}
where $\mW=[\vw_1,\dots,\vw_8]$ is the orthogonal matrix of normalized eigenvectors of $\mC$, $\Lambda=\diag{\lambda_1,\dots,\lambda_8}$ is the corresponding diagonal matrix of eigenvalues arranged in decreasing order, and $\nabla R_0(\vx)$ is a column vector. In particular, the eigenpairs satisfy the relationship
\begin{equation}
\label{eq:eigenval}
\lambda_i \;=\; \int (\nabla R_0^T \vw_i)^2\,\rho\,d\vx.
\end{equation}
Notice that if $\lambda_i=0$ for some $i \in \{1,\ldots,8 \}$, then $R_0$ is constant along the direction $\vw_i$, and therefore no change in this model output occurs from variations along the $\vw_i$ direction of the parameter input space. Such structure can be exploited to reduce the dimension of the model.
For instance, suppose $\lambda_{n+1}\ll \lambda_n$ for some $n <7$. Then, we partition the orthogonal decomposition of $\mC$ into
\begin{equation}
\label{eq:partition}
\Lambda = \bmat{\Lambda_1 & \\ & \Lambda_2},\qquad
\mW = \bmat{\mW_1 & \mW_2},
\end{equation}
where $\Lambda_1$ is a diagonal matrix of the first $n$ eigenvalues, and $\mW_1$ is the $8\times n$ matrix containing the first $n$ eigenvectors. To identify an active subspace, we must estimate the eigenpairs $\Lambda$, $\mW$ and identify a gap amongst the eigenvalues. Because the dimension of our parameter space is relatively small we use high order numerical integration rules to estimate $\mC$ in \eqref{eq:C}. More specifically, we use tensor product Gauss-Legendre quadrature on $8^8$ points (eight per parameter dimension). 

Next, define $\alpha\in\mathbb{R}^8$ by
\begin{equation}
\label{eq:actscore}
\alpha \;=\; \alpha(n) \;=\; \sum_{j=1}^n \lambda_j\,\vw_j^2,
\end{equation}
where the exponent on the vector $\vw_j$ denotes an elementwise exponentiation, i.e. within each component of $\vw_j$. The entries of the vector $\alpha(n)$ are referred to as the \emph{activity scores} of $\mC$, and we use these numbers to rank the importance of the input parameters for the quantity of interest $R_0$. The active subspace's construction provides insight into the interpretation of the activity scores \cite[Chapter~3]{constantine2015active}. Namely, the eigenvector $\vw_1$ identifies the most important direction in the parameter space in the following sense: perturbing $\vx$ along $\vw_1$ changes $R_0$ more, on average, than perturbing $\vx$ orthogonal to $\vw_1$, as demonstrated in \eqref{eq:eigenval}. The components of $\vw_1$ measure the relative change in each component of $\vx$ along this most important direction, so they impart significance to each component of $\vx$. The second most important direction is the eigenvector $\vw_2$, and the relative importance of $\vw_2$ is measured by the difference between the eigenvalues $\lambda_1$ and $\lambda_2$. For example, if $\lambda_1\gg\lambda_2$, then $R_0(\vx)$ possesses a dominant one-dimensional active subspace, and the importance of the components of $\vx$ is captured by the components of $\vw_1$. Therefore, to construct this global sensitivity metric, it is reasonable to scale each eigenvector by its corresponding eigenvalue.  

The results of computing the eigenvalues $\lambda_i$ ($i=1,...,8$) and pertinent eigenvector $\vw_1$ for $\mC$ are presented in Figures~\ref{fig:as_eigs} and \ref{fig:as_evec}, respectively.  Notice that eigenvalues are displayed on a log scale, and thus a notable (nearly two orders of magnitude) spectral gap occurs between $\lambda_1$ and $\lambda_2$ for each nation.  Hence, $\vw_1$ displays the parameters of greatest global importance to $R_0$, and we see that it is $\psi$ which is responsible for the greatest decrease in this output variable within either country.
This further justifies the omission of variations in other parameters within the local analysis of previous sections since, on average, they influence $R_0$ much less than $\psi$.
As an additional benefit of the active subspace method, we note that the computed activity score is actually independent of the values of parameters estimated from the data, e.g. $\omega_0$ and $\psi_0$.  Instead, we merely require a suitable range of parameter values in order to implement the method, and this can be much easier to obtain than a realistic fit of parameters from case data which tracks infected and deceased individuals.  As noted within Table~\ref{tab:intervals} the intervals for these parameter values are easily obtained as rate coefficients from compiled statistics \cite{NEJM}, such as the mean time from infection to recovery or infection to death.

Revisiting Figure~\ref{fig:as_eigs}, a logical argument can be made that an even larger spectral gap occurs within $\lambda_2$ and $\lambda_3$ for each nation. Thus, we have computed the corresponding activity scores $\alpha(2)$ for each country and the resulting vector is shown in Figure~\ref{fig:act_scores}.
Here, we see that the activity score for $\psi$ within both $\alpha(1)$ and $\alpha(2)$ is significantly greater within Sierra Leone than in Liberia.
This result suggests that if resources can be allocated so that the hospitalization rate within either country is altered enough to render its current (or fitted) value insignificant, then the basic reproduction number would experience a greater decrease by increasing the hospitalization rate in Sierra Leone rather than Liberia.
Thus, the benefit of the active subspace method becomes clear - if large variations of model input parameters are realistic, then the global sensitivity metric provided by this method serves as a better decision-making tool than a classical, derivative-based local sensitivity approach.
In the former situation, the analysis suggests that allocating more resources for hospitalization within Sierra Leone would have a larger impact than doing so within Liberia.

\section{Conclusions and Future Work}
\noindent To gain insight into the spread of the most recent Ebola epidemic within Western Africa, a modified SEIR model was constructed that incorporates the effects of interaction amongst infectious individuals, including those who have succumb to the disease, and the population of susceptible individuals. Upon creating the model, parameter values were fit to known WHO case data \cite{whositrep} gathered between March and December 2014 for Guinea, Liberia, and Sierra Leone. Local metrics were then proposed to determine the optimal location for the allocation of additional medical resources. Finally, a novel global sensitivity metric was proposed and explored to allow for a greater understanding of model output responses to changes within input parameters.
The former analysis clearly established that Liberia would most benefit, via a reduced $R_0$, from an increase in the hospitalization rate.
All of the software used to develop these tools has been compiled at
\begin{center}
\texttt{https://github.com/PaulMDiaz/Ebola}
\end{center}
and is freely available to any parties wishing to further the results of the article. 

Given the results of the previous sections, one must conclude that the optimal placement for treatment facilities in Western Africa depends largely on the amount of new resources that can be devoted to a specific country.
If resources are scarce and parameters cannot deviate significantly from their current values, then local metrics imply that Liberia will experience the greatest benefit from additional resources.
This can be seen in many ways, as Liberia possesses the larger current value of $R_0$, the smaller proportion of hospitalized cases, the greater rate of decrease in $R_0(\psi)$, and the greatest decrease in the infected population as $\psi$ is varied.
%
%First, the parameter values regulating hospitalization rates, and death rates, $\rho_1, \rho_2$, are the largest and most deadly in Liberia. Second, a major decrease in the infected population can be seen in Liberia when the hospitalization rate, $\psi$, varies around the steady state. Third, Liberia experiences the greatest decrease in the infected population at the end of our simulations when people are hospitalized faster ($\psi$ is increased). Finally, we see from figures \ref{fig:as_evec} and \ref{fig:act_scores} that in both Liberia and Sierra Leone the parameter $\psi$ has the largest impact on reducing $R_0$. Furthermore, Figure \ref{fig:act_scores} confirms our finding that the parameter $\omega$ has very little influence on $R_0$, but additionally we see that the parameters $\beta_2$ and $\gamma_1$ do not significantly affect the value of the reproductive ratio.  \par
%
Contrastingly, if the introduction of new resources can drastically alter parameters within the model, then the global activity scores are the pertinent metric, and they indicate that Sierra Leone will derive the greater benefit. 
In either case, our methods have clearly demonstrated the importance of the hospitalization rate amongst all other parameters which appear in the model. Namely, Figures \ref{fig:as_evec} and \ref{fig:act_scores} show that the parameter $\psi$ has the largest impact on reducing $R_0$ in both Liberia and Sierra Leone. Furthermore, Figure \ref{fig:act_scores} confirms our initial finding that the parameter $\omega$, as well as $\beta_2$ and $\gamma_1$, has very little influence on the basic reproduction number $R_0$.  \par

Given the scarce resources available for treatment in Western Africa, the conclusions drawn herein are further verified by the state of the Ebola epidemic in Liberia as of April 2015. In December of 2014, The People's Republic of China decided to place new treatment facilities within Liberia \cite{msfebolanow}, and the result has been a vast improvement in controlling the spread of the infection. As of April 29, 2015 there had been no new cases in the nation of Liberia for 34 days and many of the additional Ebola treatment centers within this region are either scheduled to scale-down, in the process of closing, or have already closed. Furthermore, the Chinese government's decision to allocate resources toward the opening of an additional Ebola treatment facility in Liberia lends credence to our findings, as the presented model and local metrics were able to arrive at the same conclusions. Given a current and detailed data set, the tools developed within the current study, including the modified SEIR model, the process for parameter estimation, and both local and global metrics for determining the need for hospital resources, could all be used to identify optimal resource allocation strategies for any number of future epidemics as well.

\section*{Acknowledgments} The authors would like to thank Sara Del Valle in the Energy and Infrastructure Analysis group at Los Alamos National Laboratory for working with the PIC Math program, suggesting the problem of interest, and providing helpful discussions and advice.

\appendix
\renewcommand{\theequation}{A.\arabic{equation}}
\section*{Appendix: Steady States and Basic Reproduction Number} \label{App:AppendixA}
\noindent
To determine any non-zero steady state solutions, the rates of change for the $S,E,I,H,$ and $R_I$ populations were set to zero. As previously described, the equations for the removed-buried and removed-recovered populations decouple from the model and thus were omitted. With this, we find
\begin{equation} \left.
\begin{aligned} 
\frac{dS}{dt} &= -S (\beta_1 I + \beta_2 R_I + \beta_3H) = 0\\
\frac{dE}{dt} &=  \beta_1 S I +\beta_2 S R_I +\beta_3SH- \delta E = 0\\
\end{aligned} \label{stdy_state1}
\right \}
 \end{equation} \\ 
and
\begin{equation}
\left.
\begin{aligned} 
\frac{dI}{dt} &=  \delta E - \gamma_1 I-\psi I = 0 \\
\frac{dH}{dt} &= \psi I - \gamma_2 H = 0\\
\frac{dR_I}{dt} &= \rho_1\gamma_1 I - \omega R_I = 0. 
\end{aligned} \label{stdy_state2}
\right \}
\end{equation}
The three equations within \eqref{stdy_state2} then imply
$$E = \frac{\gamma_1 + \psi}{\delta}I, \qquad H = \frac{\psi}{\gamma_2}I,  \qquad  R_I = \frac{\rho_1 \gamma_1}{\omega}I,$$
 so that substitution of these expressions into (\ref{stdy_state1}) yields
 \begin{equation}
 \left.
 \begin{aligned}
SI \left(\beta_1 + \frac{\beta_2 \rho_1 \gamma}{\omega} + \frac{\beta_3 \psi}{\gamma_2} \right) & = 0 \\
 I\left[-(\gamma_1 +\psi) +S\left(\beta_1 + \frac{\beta_2 \rho_1 \gamma_1}{\omega} +\frac{\beta_3 \psi}{\gamma_2} \right)  \right] & = 0.\\
 \end{aligned} \label{stdy_state3}
 \right \}
 \end{equation}
Notice that within (\ref{stdy_state3}), $S=0$ is one possible solution, which implies $I=0$ and that all populations are zero. As we assume throughout that the total population satisfies $N > 0$, this steady state is not possible.  Instead, imposing $I = 0$ within (\ref{stdy_state3}), we see that $S$ is arbitrary. However, to satisfy the total population constraint, we must have $S = 1$.  Hence, we find the unique infection-free steady state $(\bar{S},\bar{E},\bar{I},\bar{H},\bar{R}_I ) = (1,0,0,0,0)$.

To compute the basic reproduction number of the system with respect to this steady state, we employ the Next Generation Matrix method \cite{NextGen}.  In particular, computing the gains and losses matrices associated with \eqref{std_seir}, we find

$$F = \left( \begin{array}{cccc}
0      & \beta_1 \bar{S}  & \beta_3 \bar{S} & \beta_2 \bar{S} \\
0      & 			0     & 0 			   & 0			      \\
0      & 			0     & 0 			   & 0 				  \\
0	   &			0	  & 0 			   & 0		    	  \end{array} \right)$$ 
and
$$ V = \left( \begin{array}{cccc}
\delta  & 0			      & 0				  & 0 		 \\
-\delta & \gamma_1+\psi   & 0				  & 0 		 \\
0	    & -\psi 		  & \gamma_2		  & 0		 \\ 
0		& -\rho_1 \gamma_1& 0				  & \omega	 \end{array} \right).$$
Here, $F$ has been evaluated at the steady state determined above.
Inverting $V$ yields
$$ V^{-1} = \left( \begin{array}{cccc}
\frac{1}{\delta}  	\vspace{.1 cm}			  & 0			      & 0				  & 0 		 \\
\frac{1}{\gamma_1+\psi} 					  & \frac{1}{\gamma_1+\psi}   & 0	      & 0 		 \\
\frac{\psi}{\gamma_2 (\gamma_1 + \psi)}		  & \frac{\psi}{\gamma_2 (\gamma_1 + \psi)}& \frac{1}{\gamma_2}		  & 0		 \\ 
\frac{\rho_1 \gamma_1}{\omega(\gamma_1+\psi)} & \frac{\rho_1 \gamma_1}{\omega(\gamma_1+\psi)}& 0				  & \frac{1}{\omega}	 \end{array} \right)$$
and, using $\bar{S} = 1$, the resulting next generation matrix is 

$$ FV^{-1} = \left( \begin{array}{cccc}
\frac{\beta_1 + \frac{\beta_2 \rho_1 \gamma_1}{\omega} + \frac{\beta_3}{\gamma_2}\psi}{\gamma_1 + \psi} & \frac{\beta_1 + \frac{\beta_2 \rho_1 \gamma_1}{\omega} + \frac{\beta_3}{\gamma_2}\psi}{\gamma_1 + \psi} & \beta_3 \gamma_2 & \beta_2 \omega 		 \\
0   & 0 & 0 & 0 		 \\
0   & 0 & 0 & 0		     \\ 
0   & 0 & 0	& 0			 \end{array} \right).$$

Finally, it's easy to see that the spectral radius of this matrix is exactly $\left (FV^{-1} \right)_{1,1}$ as the remaining eigenvalues must be identically zero. Thus, we arrive at an explicit formula for the basic reproduction number, namely
$$R_0 := \rho \left (FV^{-1} \right ) = \frac{\beta_1 + \frac{\beta_2 \rho_1 \gamma_1}{\omega} + \frac{\beta_3}{\gamma_2}\psi}{\gamma_1 + \psi}.$$

\vspace{0.1in}

%\begin{minipage}{\linewidth}
%%\hspace{ 0.5 cm}
%\makebox[\linewidth]{%
%  
%  \includegraphics[keepaspectratio=true,scale=1]{SITREP_CASECOUNT_6.png}}
% \label{fig:casecount} 
% \vspace{-.75 cm}
%  \captionof{figure}{Case Counts - 29 April 2015 \cite{whositrep}}
%  
%\end{minipage}\\
%\newpage 
%\begin{minipage}{\linewidth}
%%\hspace{ 0.5 cm}
%\makebox[\linewidth]{%
%  
%  \includegraphics[keepaspectratio=true,scale=1]{SITREP_DAYS_SINCE_6.png}}
% \label{fig:days_since} 
% \vspace{-.75 cm}
%  \captionof{figure}{Days Since Last Case - 29 April 2015  \cite{whositrep}}
%  
%\end{minipage}\\
%
%\newpage 
%\begin{minipage}{\linewidth}
%%\hspace{ 0.5 cm}
%\makebox[\linewidth]{%
%  
%  \includegraphics[keepaspectratio=true,scale=1]{SITREP_ETC_Decomm_4.png}}
% \label{fig:ETC_Decomm} 
% \vspace{-.75 cm}
%  \captionof{figure}{ETC Decommission Status - 29 April 2015  \cite{whositrep}}
%  
%\end{minipage}\\

%\newpage
%
%\section*{\\Appendix B: ***} \label{App:AppendixB}
\bibliographystyle{acm}
\bibliography{Modified_SEIR_Model_12-21-15}

\renewcommand\thefigure{\arabic{figure}}    
\renewcommand\thetable{\arabic{table}}

\clearpage
                                     
\tikzstyle{line} = [draw, -latex']
\begin{figure}[t!]
\begin{center}
\begin{tikzpicture}[place/.style={circle,draw=blue!50,line width=0.4mm, fill=white},
   transition/.style={->,line width=0.4mm}, scale = 0.5]
\node[place] (s) at (-12, 0) {$S$};
\node[place] (e) at (-6, 0) {$E$};
\node[place] (i) at (0, 0) {$I$};
\node[place] (rb) at (6, 0) {$R_B$};
\node[place](ri) at (6, 8) {$R_I$};
\node[place](rr) at (6, -8) {$R_R$};
\node[place](h) at (0, -8) {$H$};

\path
    (s) edge[transition] node[above]{$\pmb{\beta_1 + \beta_2+ \beta_3}$} (e)
    (e) edge[transition] node[above]{$\pmb{\delta}$} (i)
    (i) edge[transition] node[left]{$\pmb{\rho_1\gamma_1}$} (ri)
    (ri) edge[transition] node[right]{$\pmb{\omega}$} (rb)
    (i) edge[transition] node[right, pos = 0.7]{$\pmb{(1- \rho_1)\gamma_1}$} (rr)
    (i) edge[transition] node[left]{$\pmb{\psi}$} (h)
    (h) edge[transition] node[above]{$\pmb{(1 - \rho_2)\gamma_2}$} (rr)
    (h) edge[transition] node[left, pos = 0.8]{$\pmb{\rho_2\gamma_2}$} (rb)
;
\end{tikzpicture}
\caption{ \footnotesize A graph representing the states ($S$, $E$, $I$, $R_B$, $R_I$, $R_R$, and $H$) and transition pathways (arrows) in the Ebola model \eqref{std_seir}. Table \ref{tab:paramdescrip} further describes the model parameters included above.}
\label{tikzCM}
\end{center}
\vspace{-0.1in}
\end{figure}
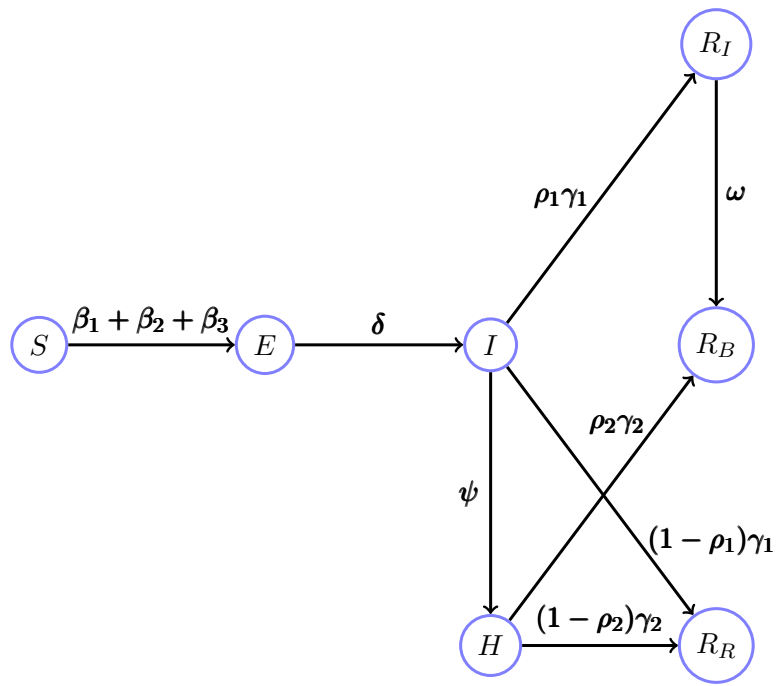

\clearpage

\begin{table}[h]
\small
\centering
\begin{tabular}{ | c | l |}
\hline 
Parameter & Description\\
\hline $\beta_1$    & transmission rate between infected and susceptible \\
\hline $\beta_2$	& transmission rate between removed but still infectious and susceptible \\
\hline $\beta_3$    & transmission rate between hospitalized and susceptible\\
\hline $\delta$   & (incubation period)$^{-1}$\\
\hline $\gamma_1$ & (average time with disease for unhospitalized individuals)$^{-1}$ \\
\hline $\gamma_2$ & (average time with disease for hospitalized individuals)$^{-1}$\\
\hline $\psi$   & (average time for infected to become hospitalized)$^{-1}$\\
\hline $\rho_1$   & proportion of infected who die of the disease and are not hospitalized \\
\hline $\rho_2$   & proportion of infected who die of the disease and are hospitalized\\
\hline $\omega$  & (average time until a deceased individual is properly buried)$^{-1}$\\
\hline
\end{tabular}
\caption{\footnotesize Parameters in the Ebola dynamics model.}
\label{tab:paramdescrip}
\end{table}

\clearpage

\newcolumntype{A}{ >{\centering\arraybackslash} m{1.2cm} }
\newcolumntype{B}{ >{\centering\arraybackslash} m{3cm} }
\newcolumntype{C}{ >{\centering\arraybackslash} m{.01cm} }

\begin{table}[t]
\small
\begin{center}
    \begin{tabular}{ | B | B | B | B | A C |}
    \hline    Parameter           & Guinea              & Liberia            	& Sierra Leone       	& Source &  \\ [1.5ex] \hline
    $\beta_1$  			 & $0.315$ 	  	& $0.376$   		& $0.251$ 		& Fit    & \\[1ex] \hline
    $\beta_2$  		 	 & $0.16$  		& $0.135$   		& $0.395$ 		& Fit    & \\[1ex] \hline
    $\beta_3$  			 & $0.0165$	       	& $0.163$ 		& $0.079$ 		& Fit   & \\[1ex] \hline
    $\delta$   			 & $\frac{1}{9}$  	& $\frac{1}{9}$  	& $\frac{1}{9}$      	&\cite{dixon2014ebola}&\\[1ex] \hline
    $\gamma_1$    		& $0.295$ 		& $0.0542$ 		& $0.051$ 	    	& Fit&\\[1ex] \hline
    $\gamma_2$ 		& 0.016         		& 0.174               	&  0.0833			& Fit	    &\\[1ex] \hline
    $\psi$   			& 0.500           		& 0.500                	&  0.442			& Fit        &\\[1ex] \hline
    $\rho_1$ 			& 0.98              		& 0.98           		&  0.76		        & Fit%, \cite{dixon2014ebola}  
    & \\ [1ex] \hline
    $\rho_2$ 			& 0.93               	& 0.88           		&  0.74			& Fit%, \cite{dixon2014ebola}  
    & \\ [1ex] \hline
    $\omega$ 			& 0.300                 	& 0.325        		&  0.370			& Fit   	    & \\ [1ex] \hline
    \end{tabular}
    \caption{\footnotesize Fitted parameter values for each country. %The values are fit with a nonlinear least squares procedure.  
    The parameter $\delta$ was not fit, but taken directly from \cite{dixon2014ebola}, while initial guesses for $\rho_1$ and $\rho_2$ were motivated by this paper.}
    \label{tab:Param}
\end{center}
\end{table}

\clearpage

\begin{figure}[t]
        \centering
        \begin{subfigure}[b]{0.52\textwidth}
                \includegraphics[width=\textwidth]{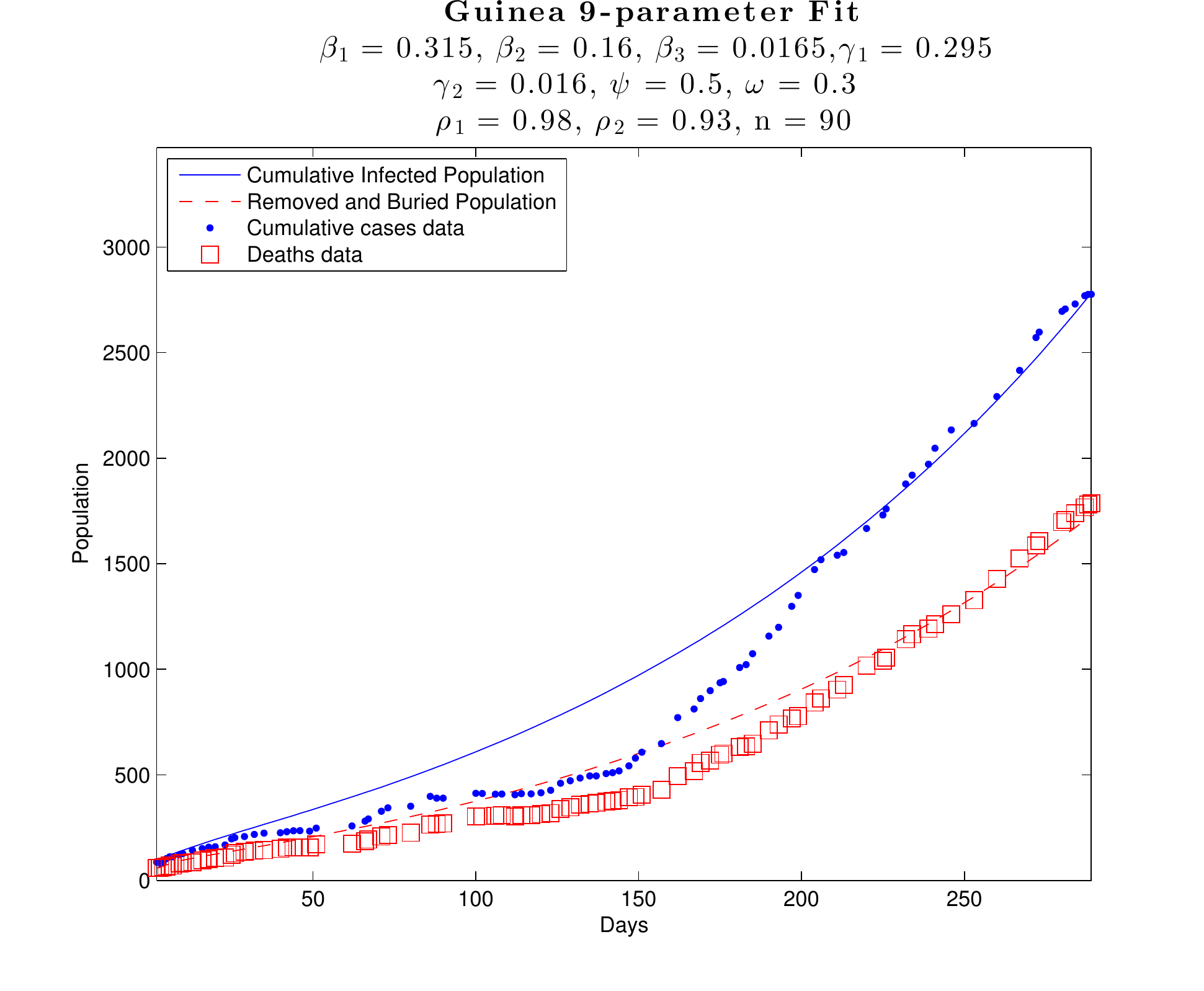} %Guinea_fit_5-23
                \vspace{-0.3in}
                \caption{\footnotesize Guinea}
        \end{subfigure}
        \hspace{-0.35in}
        \begin{subfigure}[b]{0.52\textwidth}
                \includegraphics[width=\textwidth]{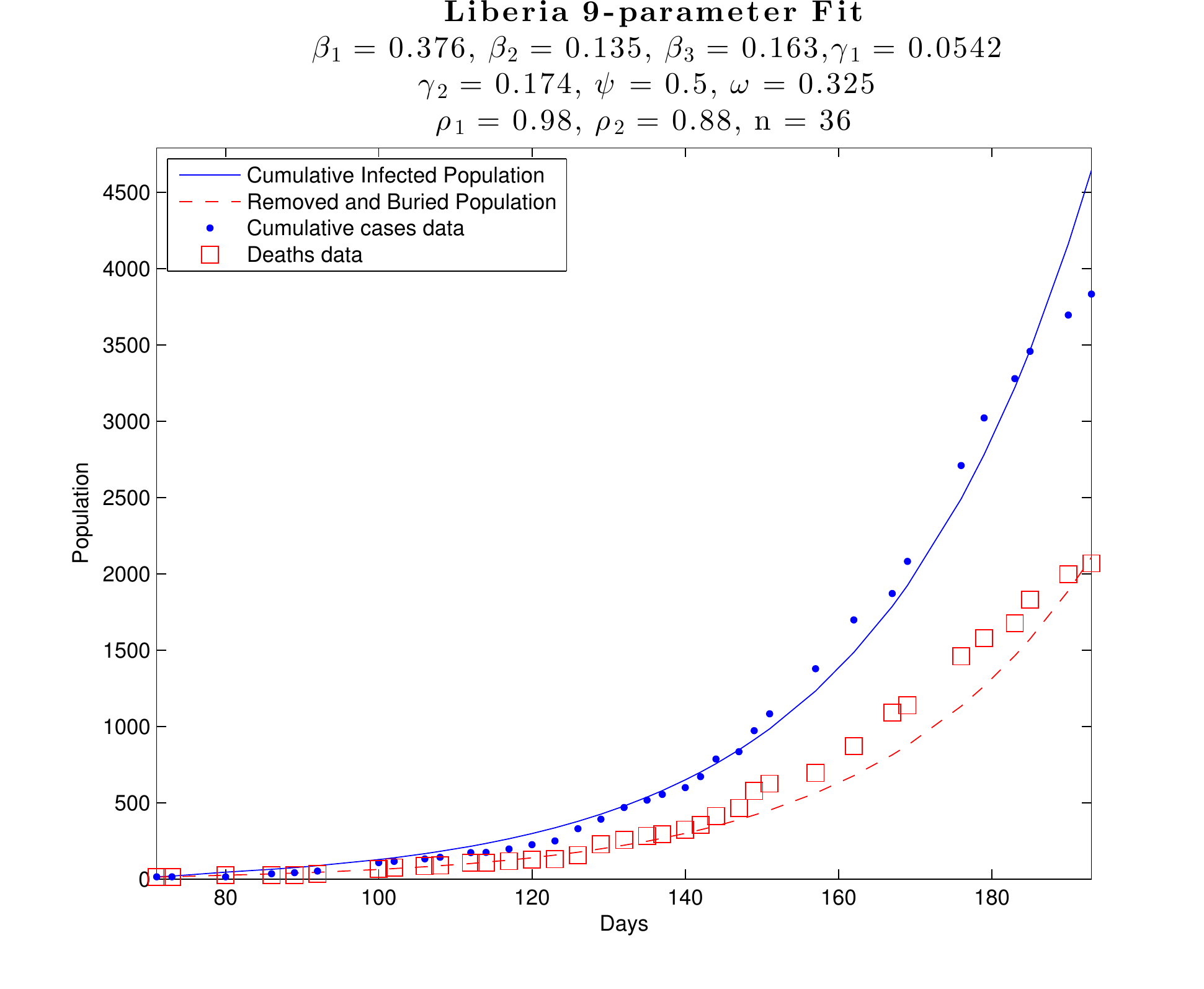} %Liberia_fit_5-23
                \vspace{-0.3in}
                \caption{\footnotesize Liberia}
        \end{subfigure}
        \\

        \begin{subfigure}[b]{0.52\textwidth}
                \includegraphics[width=\textwidth]{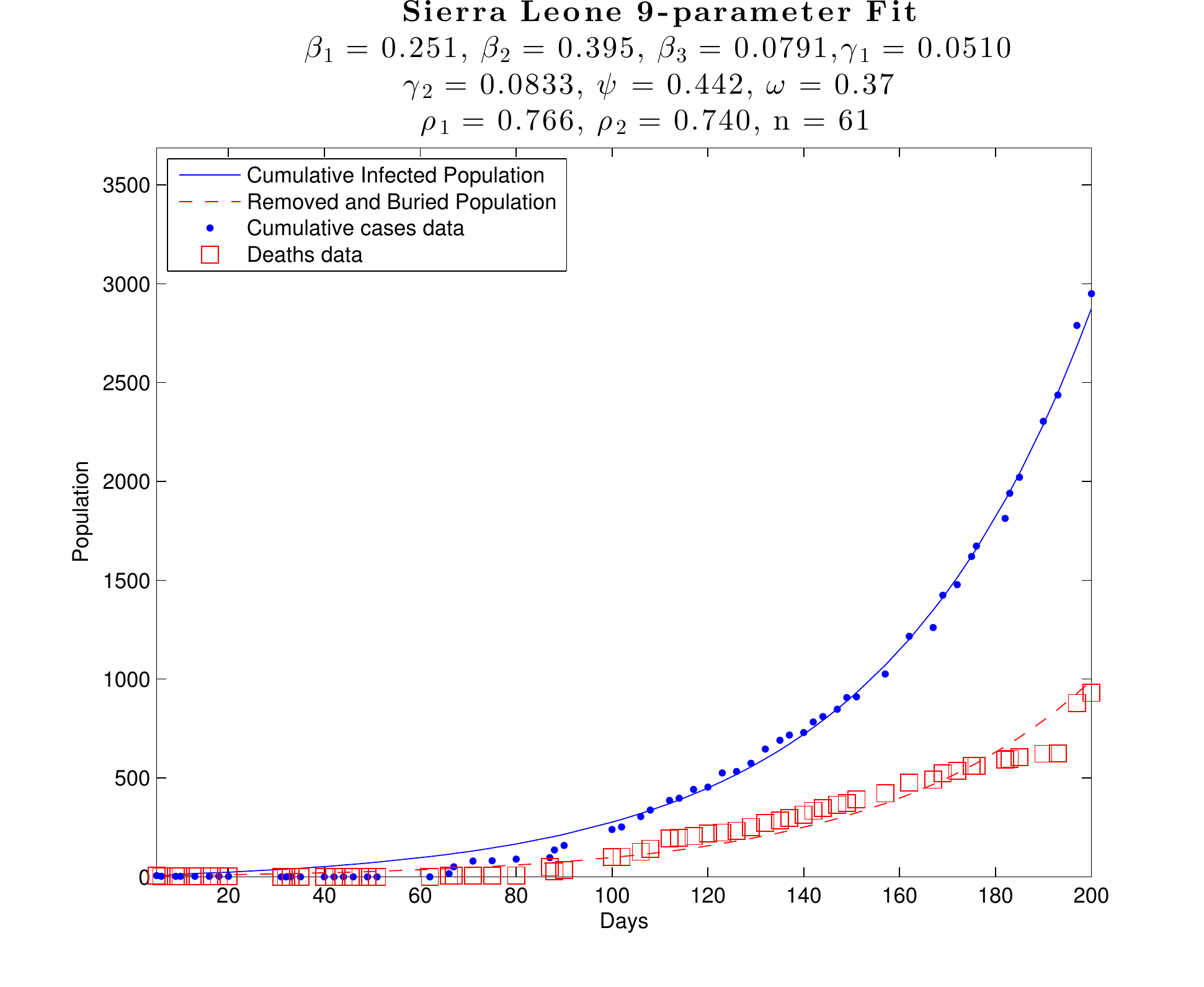} %SL_fit_5-23
                \vspace{-0.3in}
                \caption{\footnotesize Sierra Leone}
        \end{subfigure}
	\vspace{-0.1in}
        \caption{\footnotesize Comparisons of fitted models to available date for Guinea, Liberia, and Sierra Leone.}
\label{fig:paramgraphs}
\end{figure}

\clearpage

\begin{table}[t]
	\parbox{1\linewidth}{\centering 
		\begin{tabular}{| c | c | c | c |}
			\hline
			\multicolumn{4}{| c |}{\textbf{Reproduction Numbers}} \\
			\hline
			\pbox{\textwidth}{\textbf{Country}}&\pbox{\textwidth}{Guinea}&\pbox{\textwidth}{Liberia}&\pbox{\textwidth}{Sierra Leone}  \\ 
			\hline 
			\pbox{\textwidth}{$R_0$}&\pbox{\textwidth}{$1.241$}&\pbox{\textwidth}{$1.563$}&\pbox{\textwidth}{$1.445$} \\
	\hline
\end{tabular}
}
\caption{\footnotesize{Basic reproduction numbers obtained from fit parameters} \label{tab:R0}}
\end{table} 

\clearpage

\begin{table}[t]
\begin{center}
    \begin{tabular}{ | l | c | c | c | c |}
    \hline 
    Description &       Liberia  & Sierra Leone            \\  \hline
    Time to hospitalization 			        & 1.2 days      	 &  1.2 days		 	\\  \hline
    Time from hospitalization to death 		& 4.38 days	 &  10.09 days			\\  \hline
    Time from hospitalization to recovery 	& 15.80 days	 &  17.16 days			\\  \hline
    Time from infection to death, unhospitalized & 7.5 days	 &   7.6 days			\\  \hline
    Time from infection to recovery, unhospitalized & 9.34 days	 & 8.45 days		\\  \hline
    Time to proper burial 				& 3.08 days            &  2.70 days      \\ \hline
    Proportion of hospitalized cases         	& 0.57    	      &  0.60	   \\  \hline
    \end{tabular}
    \caption{\footnotesize{Relevant parameter values in determining hospital placement.  Here, the fatality rates from \cite{NEJM} have been used to compute these values, namely Liberia - $72.3\%$ (unhospitalized), $67\%$ (hospitalized) and Sierra Leone - $69\%$ (unhospitalized) and $61.4\%$ (hospitalized). }}
\label{tab:Spef}
\end{center}
\vspace{-0.2in}
\end{table}

\clearpage

\begin{figure}[t]
        \hspace{-0.2in}
        \begin{subfigure}[b]{0.55\textwidth}
                \includegraphics[width=\textwidth]{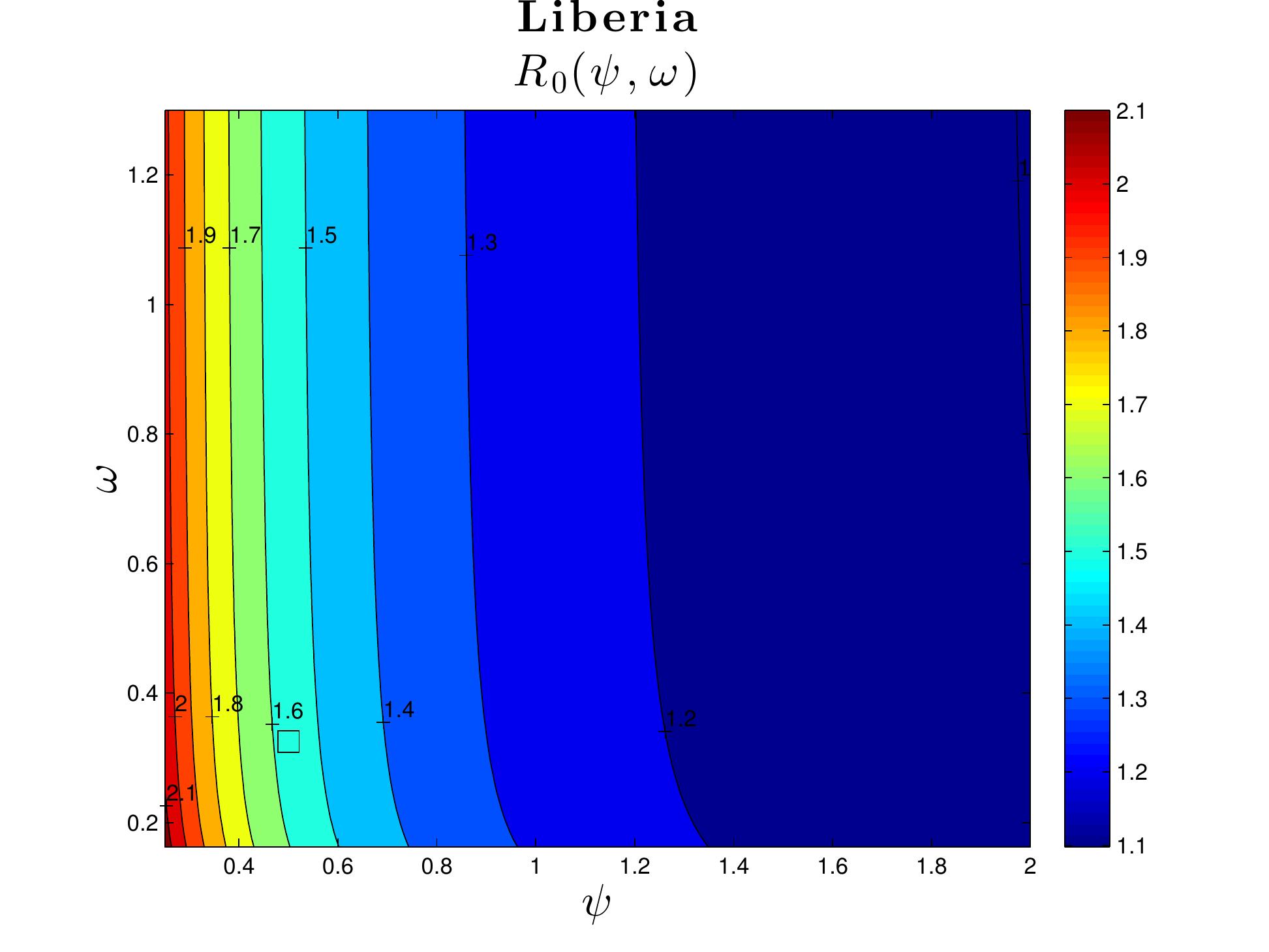}  %Liberia_fit_5-23
                \vspace{-0.3in}
                %\caption{\footnotesize Liberia}
        \end{subfigure}
        \hspace{-0.3in}
        \begin{subfigure}[b]{0.55\textwidth}
                \includegraphics[width=\textwidth]{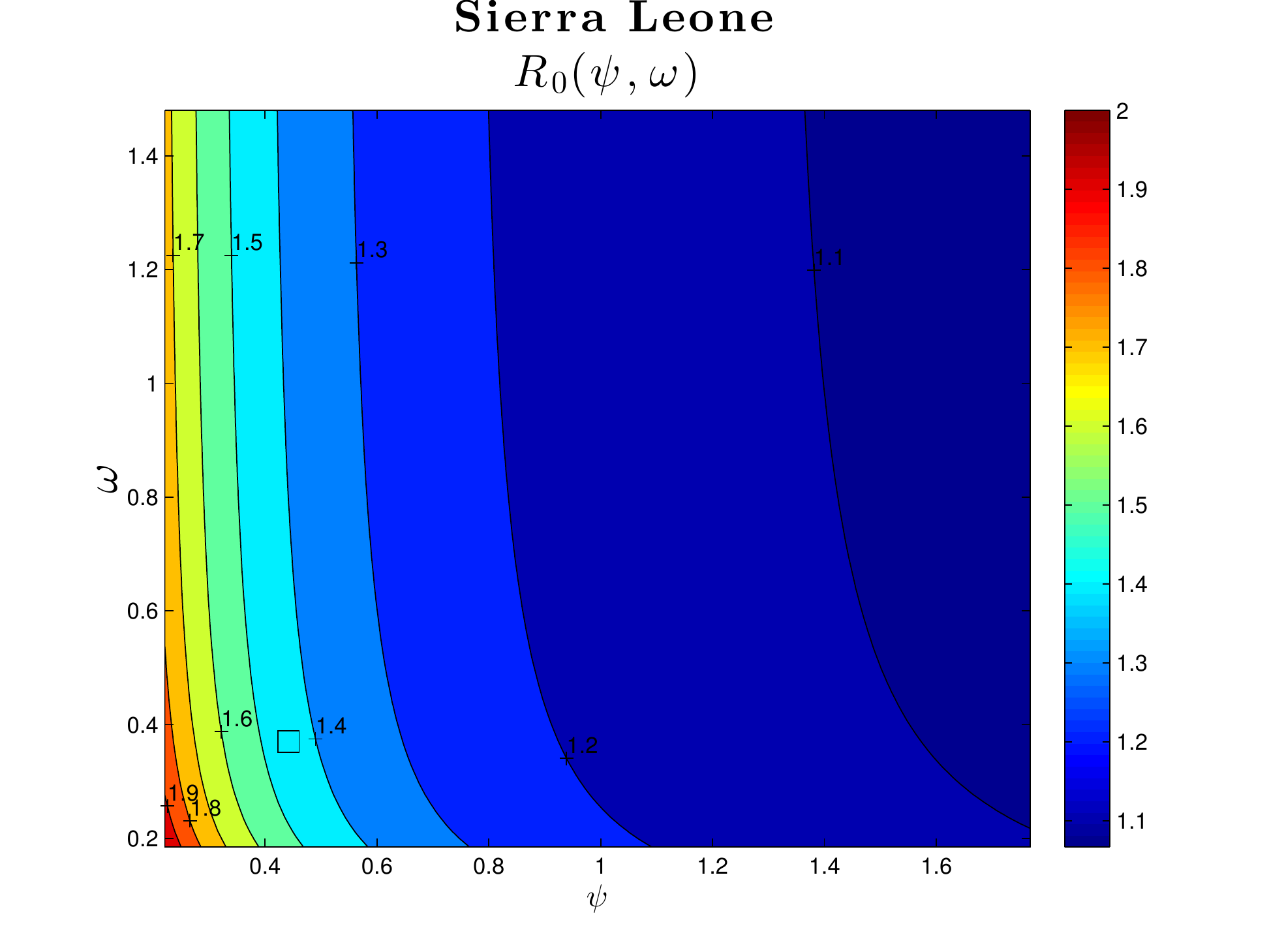} %SL_fit_5-23
                \vspace{-0.3in}
                %\caption{\footnotesize Sierra Leone}
        \end{subfigure}
\caption{\footnotesize{Contour plots of $R_0(\psi,\omega)$.  Parameters were varied from one-half their fitted value to four times this number. The black square represents the fit parameter pair $(\psi_0, \omega_0)$.} \label{fig:contours}}
\end{figure}

\clearpage

\begin{table}
\parbox{.5\linewidth}{
\centering
\begin{tabular}{| c | c | c |}
			\hline
			\multicolumn{3}{| c |}{\textbf{Sensitivities of Reproduction Number}} \\
			\hline
			\pbox{\textwidth}{\textbf{Country}}&\pbox{\textwidth}{Liberia}&\pbox{\textwidth}{Sierra Leone}  \\ 
			\hline 
			\pbox{\textwidth}{\ \\ $\displaystyle \Upsilon^{R_0}_\psi (\psi_0,\omega_0)$ \\} &\pbox{\textwidth}{$-0.3616$}&\pbox{\textwidth}{$-0.3074$} \\
			\hline
			\pbox{\textwidth}{\ \\ $\displaystyle \Upsilon^{R_0}_\omega (\psi_0,\omega_0)$ \\}&\pbox{\textwidth}{$-0.0255$}&\pbox{\textwidth}{$-0.0585$} \\
%			\hline
%			\pbox{\textwidth}{\ \\ $\displaystyle \vert \nabla R_0(\psi_0, \omega_0) \vert$ \\}&\pbox{\textwidth}{$1.137$}&\pbox{\textwidth}{$1.031$} \\
			
	\hline
\end{tabular}
\caption{\footnotesize{Changes in $\Upsilon^{R_0}(\psi,\omega)$ at fitted values} 
\label{tab:R0prime_2param}}
}
\hfill
\parbox{.5\linewidth}{
\centering
\begin{tabular}{| c | c | c |}
			\hline
			\multicolumn{3}{| c |}{\textbf{Marginal Change in $R_0$}} \\
			\hline
			\pbox{\textwidth}{\textbf{Country}}&\pbox{\textwidth}{Liberia}&\pbox{\textwidth}{Sierra Leone}  \\ 
			\hline 
			\pbox{\textwidth}{\ \\ $\displaystyle R_0'(\psi_0)$\\ }&\pbox{\textwidth}{$-1.131$}&\pbox{\textwidth}{$-1.005$} \\
			\hline
			\pbox{\textwidth}{\ \\ $\displaystyle \inf_{\psi} R_0(\psi)$ \\ }&\pbox{\textwidth}{$0.937$}&\pbox{\textwidth}{$0.948$} \\
			\hline
			\pbox{\textwidth}{\ \\ $\displaystyle \sup_{\psi} R_0(\psi)$ \\}&\pbox{\textwidth}{$7.344$}&\pbox{\textwidth}{$5.740$} \\ \hline
\end{tabular}
\caption{\footnotesize{Changes in $R_0(\psi)$ at differing $\psi$} \label{tab:R0prime}}
}
\end{table}

\clearpage

\begin{figure}[!t]
        %\centering
        \hspace{-0.4in}
        \begin{subfigure}[b]{0.48\textwidth}
                \includegraphics[width=1.2\textwidth]{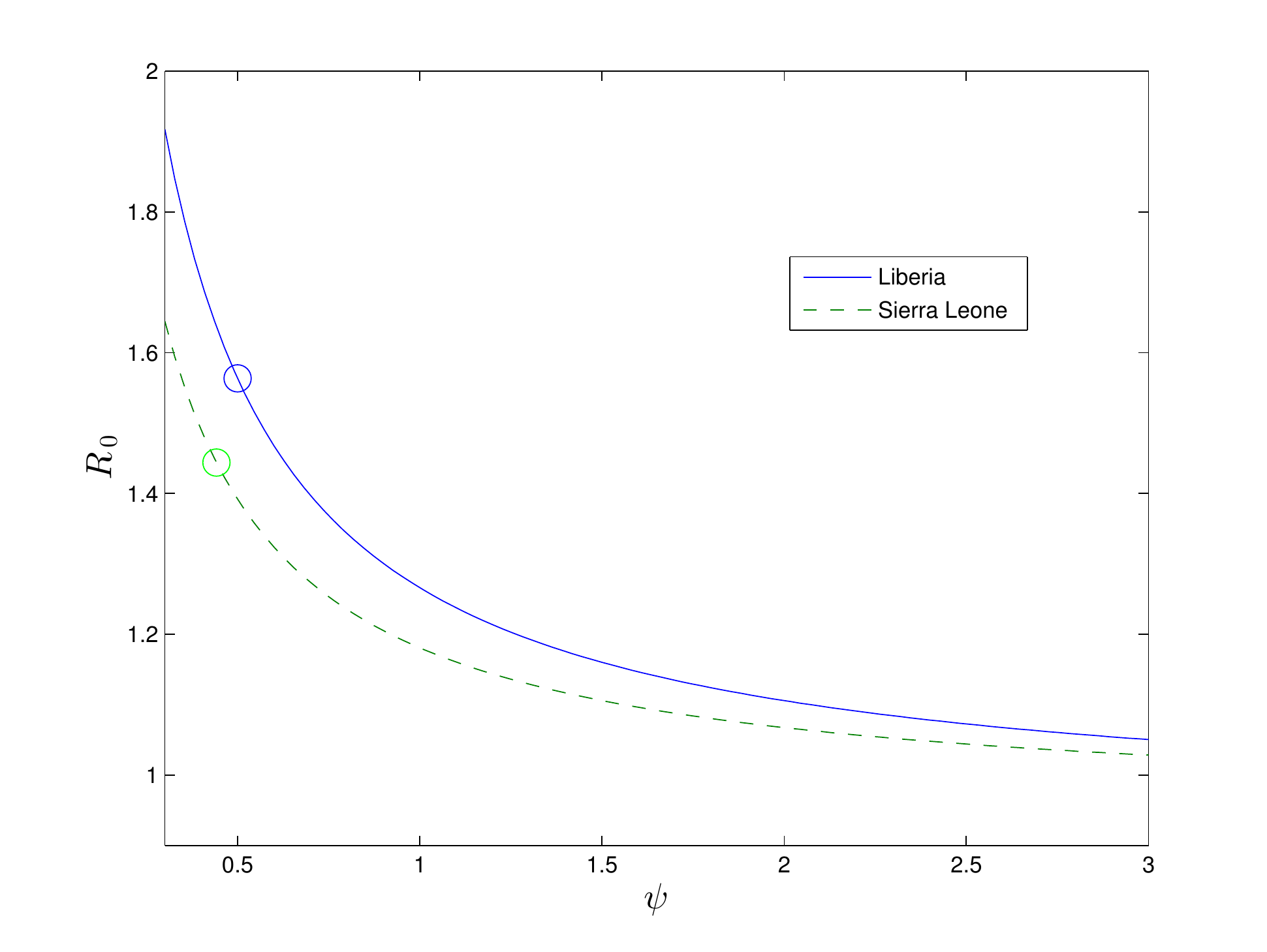} %R0plot_10-15
                \vspace{-0.3in}
                %\caption{}
        \end{subfigure}
        \hspace{0.2in}
        \begin{subfigure}[b]{0.48\textwidth}
                \includegraphics[width=1.2\textwidth]{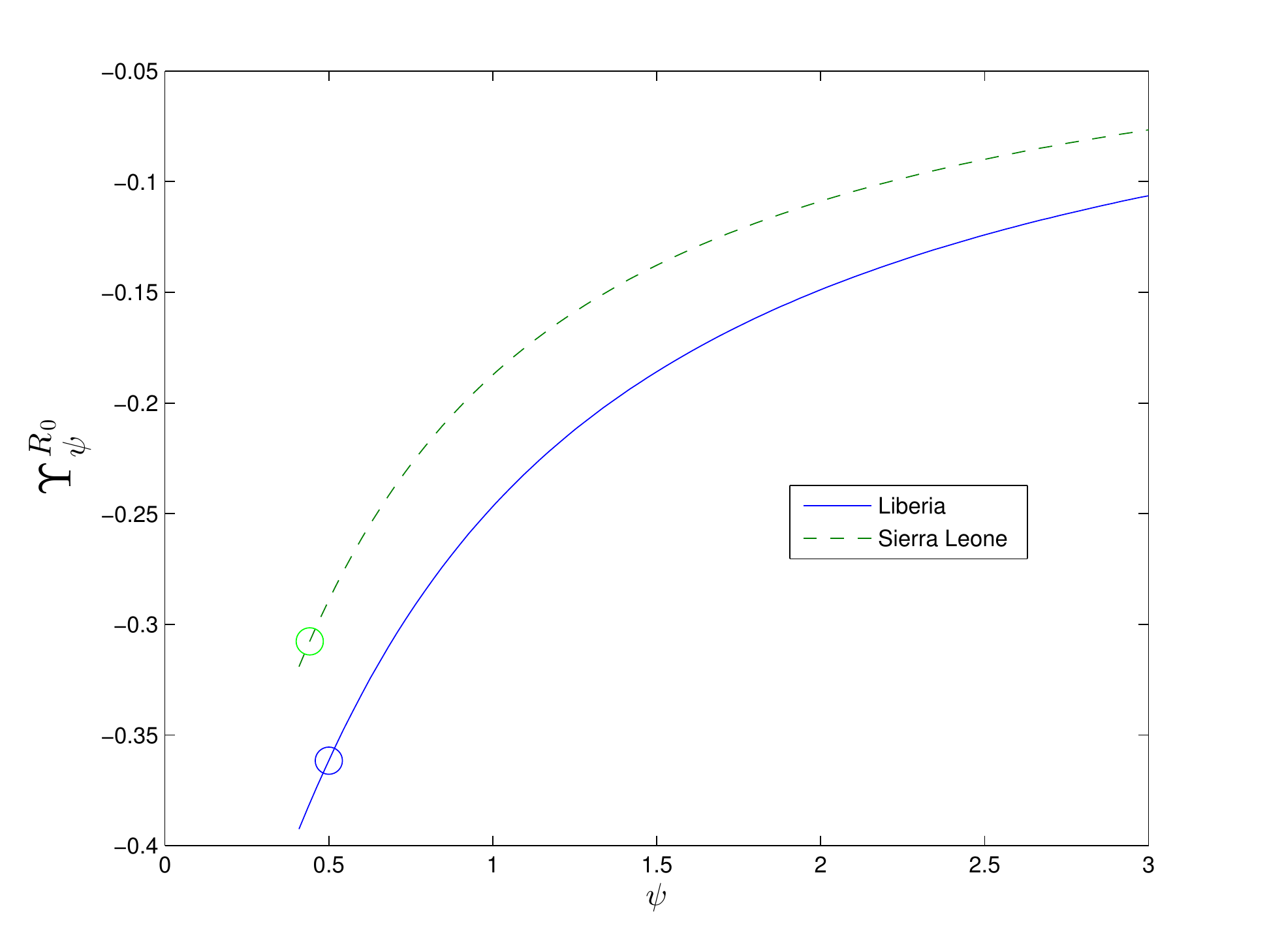} %R0primeplot_10-15 %R0primeplot_1-4
                \vspace{-0.3in}
                %\caption{Derivative of $R_0$ with respect to hospitalization parameter}
        \end{subfigure}
\caption{\footnotesize{Reproduction numbers by nation. $R_0$ as a function of hospitalization parameter (left) and
the normalized forward sensitivity index of $R_0$ with respect to hospitalization parameter (right). 
The circles represent the values of these quantities at the hospitalization level obtained from data.} \label{fig:R0graphs}}
\end{figure}

\clearpage

\begin{figure}[t]
        \centering
        \begin{subfigure}[b]{0.52\textwidth}
                \includegraphics[width=\textwidth]{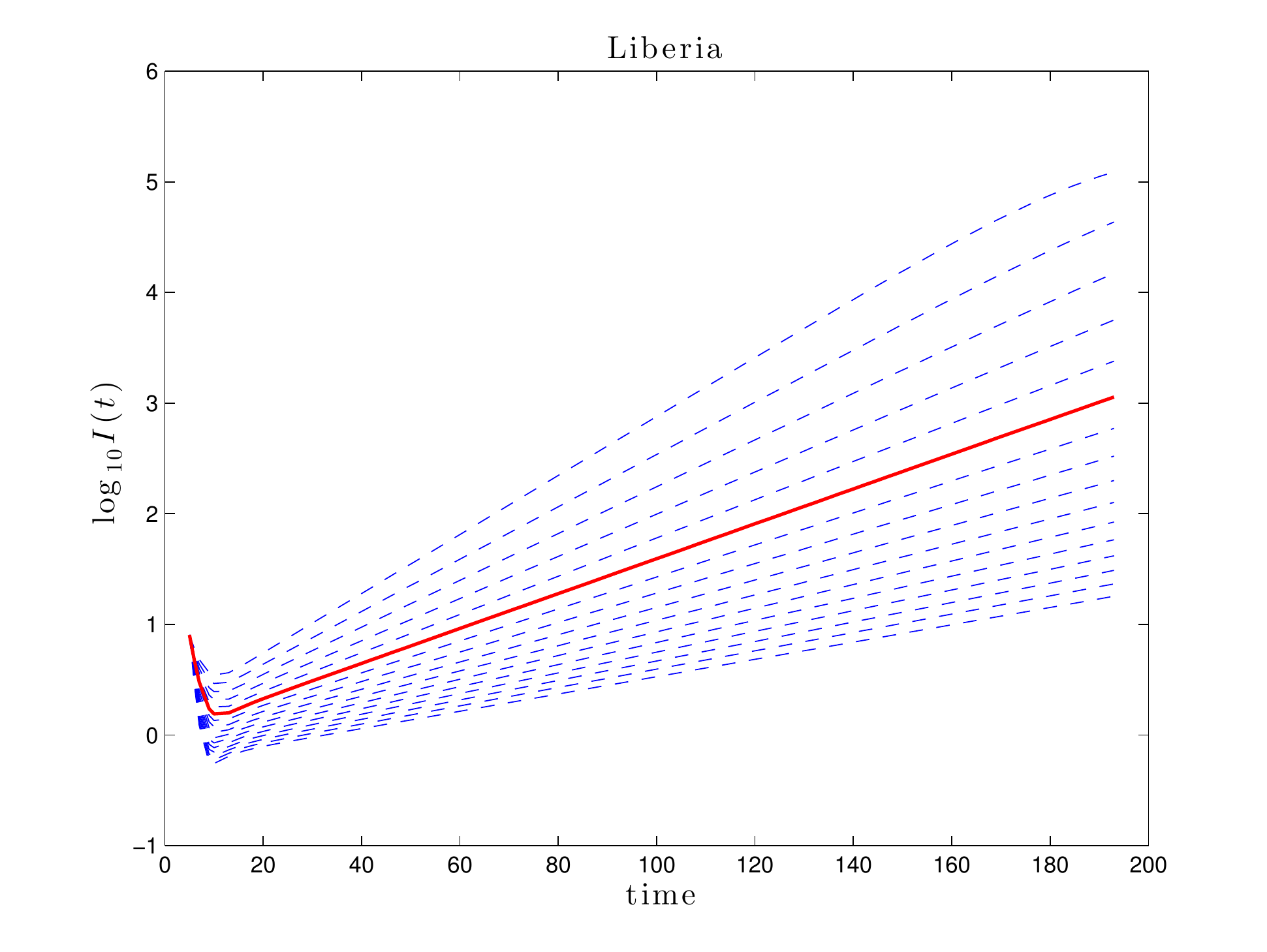} 
                \vspace{-0.3in}     
                \caption{\footnotesize Liberia}     
        \end{subfigure}
        \hspace{-0.35in}
        \begin{subfigure}[b]{0.52\textwidth}
                \includegraphics[width=\textwidth]{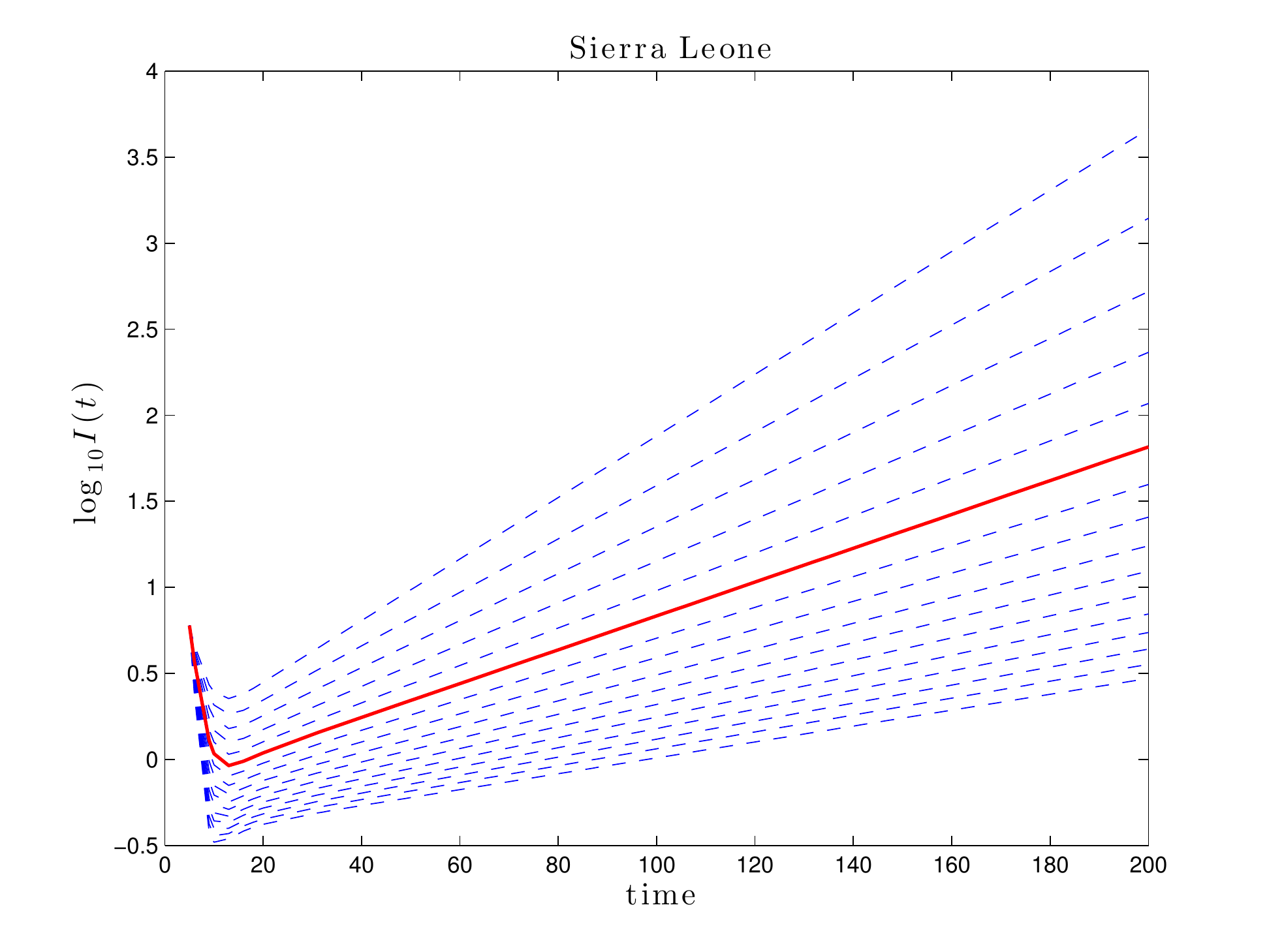} 
                \vspace{-0.3in}
                \caption{\footnotesize Sierra Leone}
        \end{subfigure}
        \caption{\footnotesize{Local sensitivity analysis of the infected population $I(t)$ (on a log-scale) as a function of the hospitalization parameter $\psi$. This parameter was varied from one-half its fitted value to twice this number. Notice the difference in scales between the nations.} \label{fig:sensitivity}}
	\vspace{-0.1in}
\end{figure}

\clearpage

\begin{figure}[ht!]
		\vspace{-1in}
        \centering
        \begin{subfigure}[b]{0.52\textwidth}
                \includegraphics[width=\textwidth]{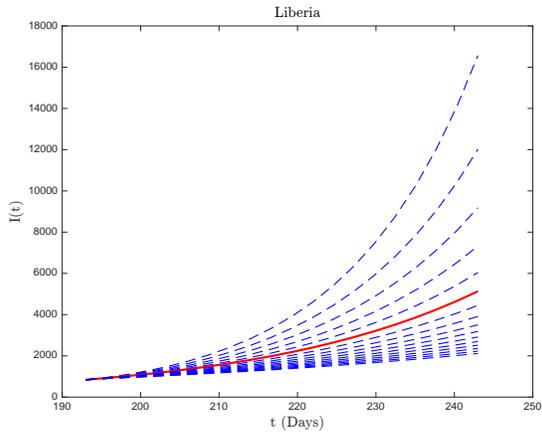} 
                \vspace{-1.2in}
                \caption{\footnotesize Liberia}
        \end{subfigure}
        \hspace{-0.35in}
        \begin{subfigure}[b]{0.52\textwidth}
                \includegraphics[width=\textwidth]{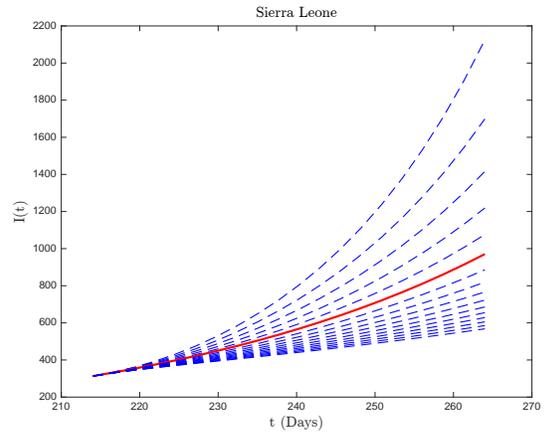} 
                \vspace{-1.2in}
                \caption{\footnotesize Sierra Leone}
        \end{subfigure}        
        \caption{\footnotesize{Future sensitivity analysis of the infected population $I(t)$ with multiplicative variation of the hospitalization parameter $\psi$. This parameter was varied from $0.5\psi_0$ to $2\psi_0$ in $0.1\psi_0$ increments, with the solid line representing the fitted value. The lines above the solid line, at the final time, represent smaller values, while the lines below represent larger values of $\psi$.} \label{fig:future_psi_sensitivity}}
      	\vspace{-0.1in}
\end{figure}
 
 \clearpage

\begin{table}[t]
\begin{center}
    \begin{tabular}{ |c | | c | c | c | c |}
    \hline 
    Parameter & Liberia       	   & Sierra Leone            \\  \hline
    $\beta_1 $ & $[0.1,0.4]$    	   &  $[0.1,0.4]$  	 	\\  \hline
    $\beta_2 $ & $[0.1,0.4]$	 	   &  $[0.1,0.4]$  			\\  \hline
    $\beta_3 $ & $[0.05,0.2]$		   &  $[0.05,0.2]$  		\\  \hline
    $\rho_1  $ & $[0.41,1]$	       &  $[0.41,1]$  		\\  \hline
    $\gamma_1 $& $[0.0276,0.1702]$&  $[0.0275,0.1569]$  	\\  \hline
    $\gamma_2 $& $[0.081,0.2100]$    &  $[0.1236,0.3840]$        \\ \hline
   	$\omega $  & $[0.25,0.5]$  	   &  $[0.25,0.5]$  	   \\  \hline
   	$\psi $    & $[0.0833,0.7]$      &  $[0.0833,0.7]$     \\  \hline
    \end{tabular}
    \caption{\footnotesize{ Intervals of parameter values for Liberia and Sierra Leone. Each of these parameters define the dimensions of the input parameter space $\Omega$. Because this is a novel model, exact intervals for each parameter are neither known nor available. The intervals for the parameters $\beta_1,\beta_2,\beta_3$, and $\omega$ come from the constraints in the barrier method for fitting parameter values in Section \ref{subsec:fits}. We used data from \cite{NEJM} to construct estimates of the remaining parameter intervals.}}
\label{tab:intervals}
\end{center}
\vspace{-0.2in}
\end{table}

\clearpage

\begin{figure}[h]
\vspace{-1in}
     \hspace{-0.2in}
        \begin{subfigure}[b]{0.55\textwidth}
                \includegraphics[width=\textwidth]{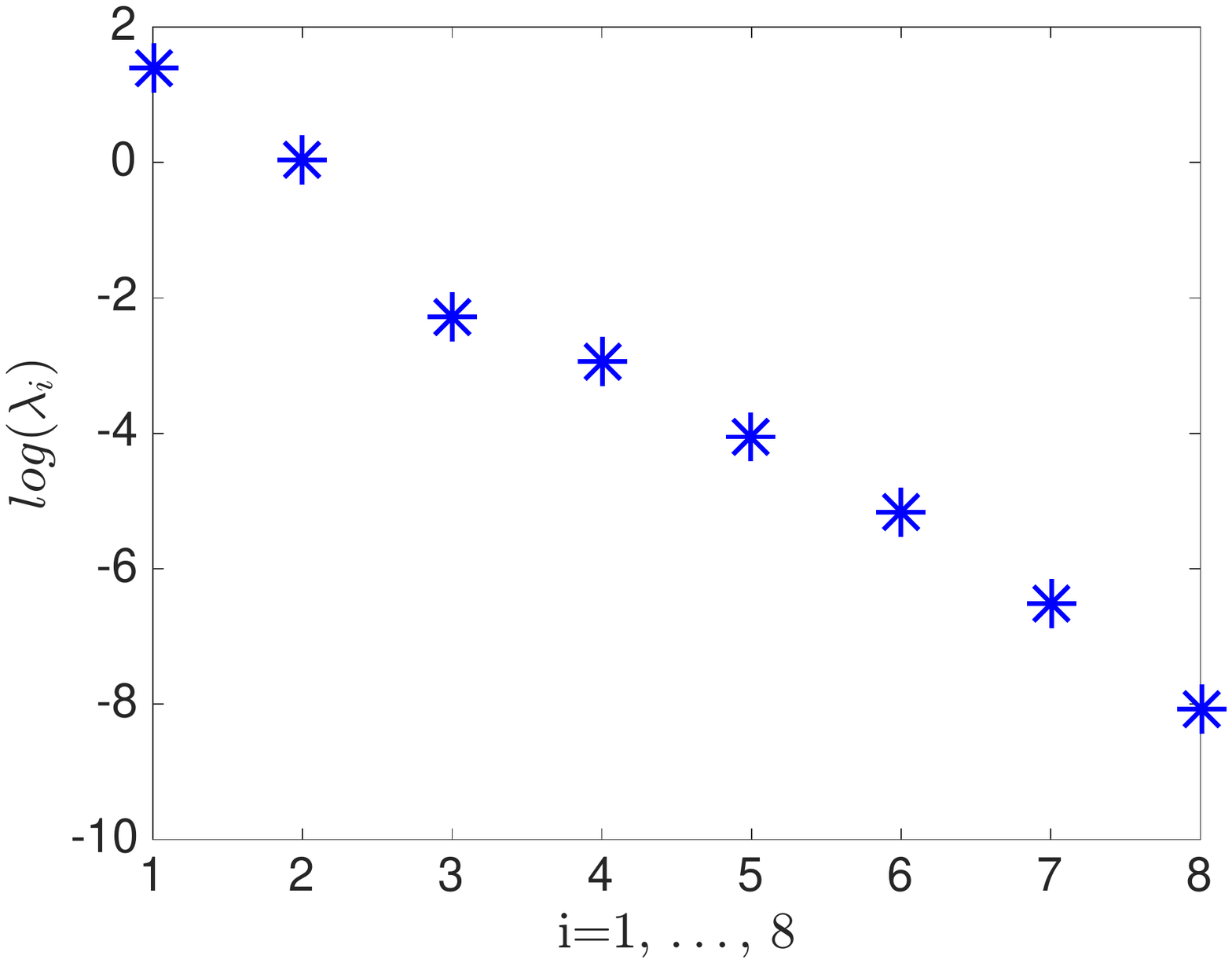} %Guinea_fit_5-23
                \vspace{-1.3in}
                \caption{\footnotesize Liberia}
        \end{subfigure}
        \hspace{-0.3in}
        \begin{subfigure}[b]{0.55\textwidth}
                \includegraphics[width=\textwidth]{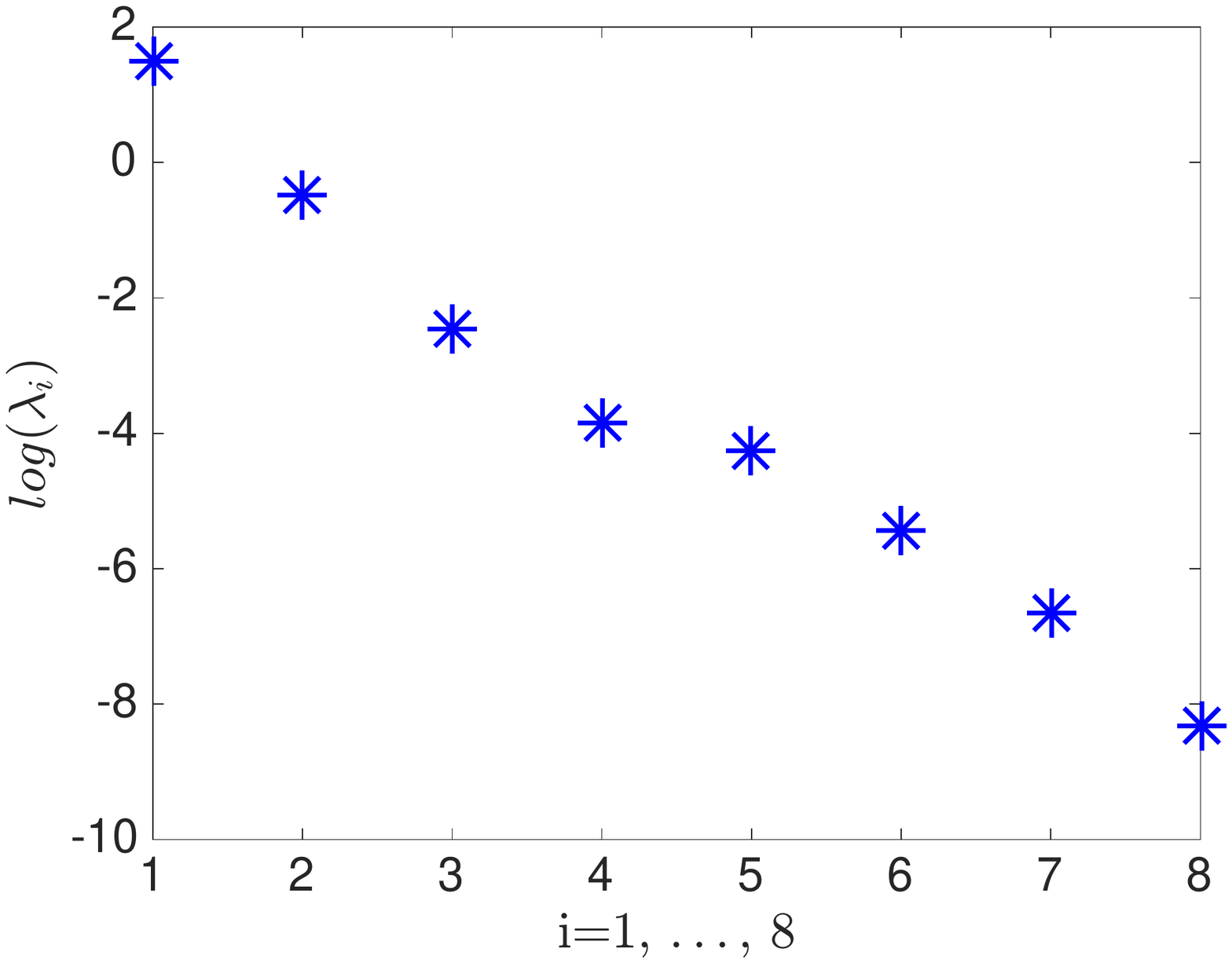} %Liberia_fit_5-23
                \vspace{-1.3in}
                \caption{\footnotesize Sierra Leone}
        \end{subfigure}
        
        \caption{\footnotesize{Magnitudes of the eigenvalues $\mLambda$ from \eqref{eq:C} on a logscale. The gap between the second and third eigenvalues for both Liberia and Sierra Leone suggests a two-dimensional active subspace, i.e., $n=2$. The computation of the matrix $\mC = \mW \mLambda \mW^T$ in \eqref{eq:C} (and therefore the eigenvalues) is done via tensor product Gauss-Legendre quadrature on $8^8$ points (eight per parameter dimension).} \label{fig:as_eigs}}
\end{figure}

\clearpage

\begin{figure}[h]
	 \vspace{-1in}
     \hspace{-0.2in}
        \begin{subfigure}[b]{0.55\textwidth}
                \includegraphics[width=\textwidth]{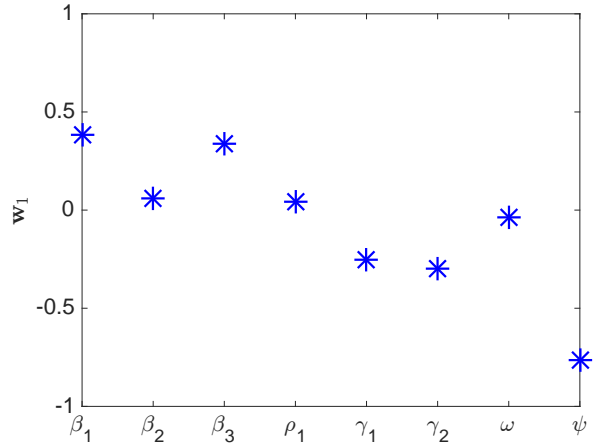} %Guinea_fit_5-23
                \vspace{-1.3in}
                \caption{\footnotesize Liberia}
        \end{subfigure}
        \hspace{-0.3in}
        \begin{subfigure}[b]{0.55\textwidth}
                \includegraphics[width=\textwidth]{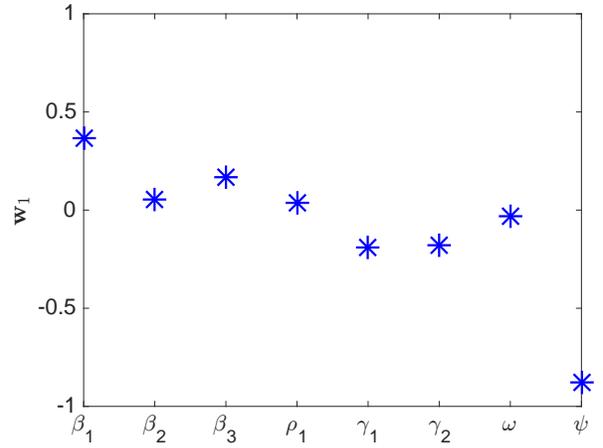} %Liberia_fit_5-23
                \vspace{-1.3in}
                \caption{\footnotesize Sierra Leone}
        \end{subfigure}
        
        \caption{\footnotesize{Components of the first eigenvector of $\mW_1$ in \eqref{eq:partition}. Each component corresponds to a different parameter as indicated. The sign of the components inform us as to how, on average, changes in the parameter correspond to changes in $R_0$. We see that the parameter corresponding to the greatest decrease in $R_0$, in both countries is $\psi$. The computation of the matrix $\mC = \mW \mLambda \mW^T$ in \eqref{eq:C} (and therefore the eigenvectors) is done via tensor product Gauss-Legendre quadrature on $8^8$ points (eight per parameter dimension).} \label{fig:as_evec}}
\end{figure}

\clearpage
        
\begin{figure}[h!]
	 \vspace{-1in}
     \hspace{-0.2in}
        \begin{subfigure}[b]{0.55\textwidth}
                \includegraphics[width=\textwidth]{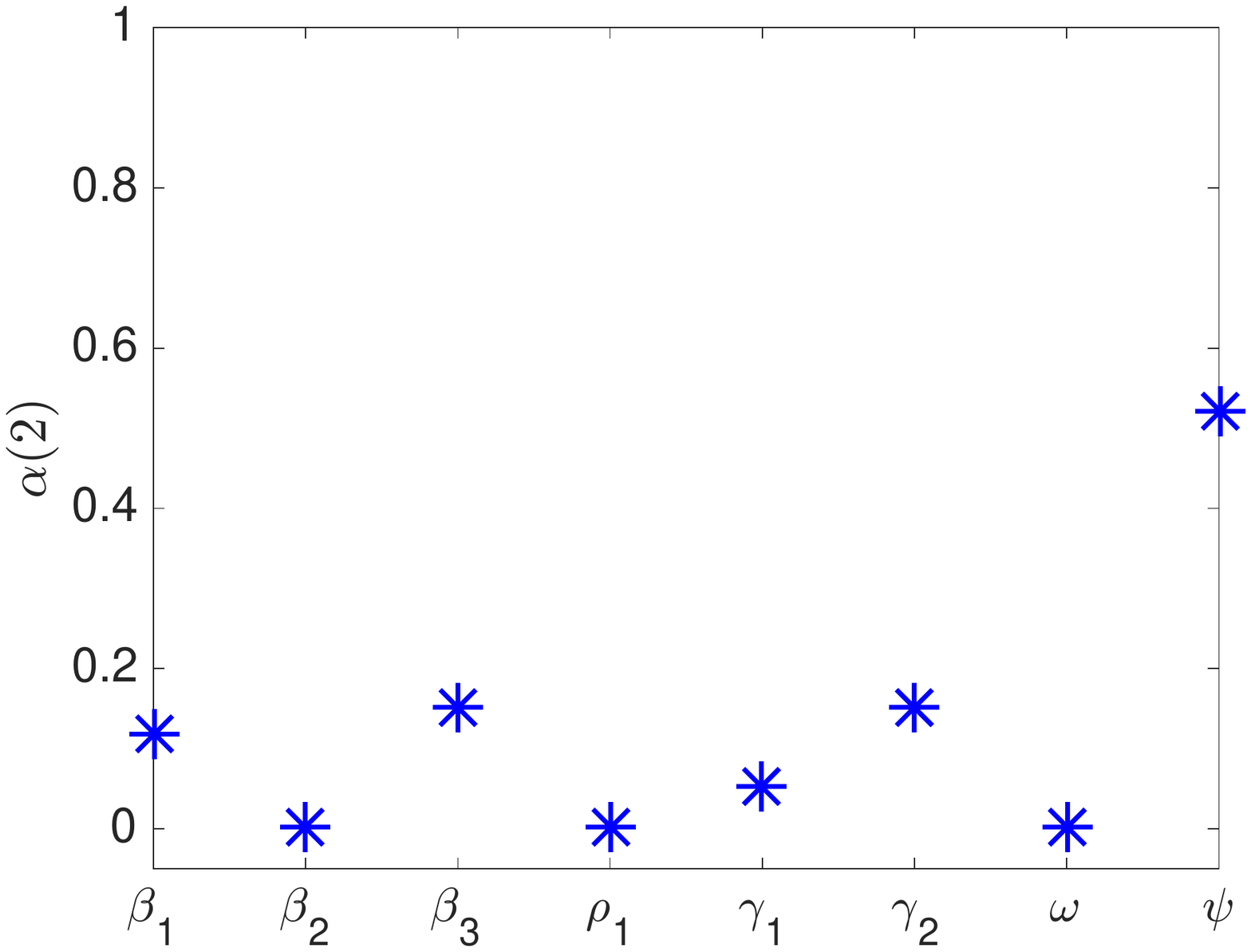} %Guinea_fit_5-23
                \vspace{-1.3in}
                \caption{\footnotesize Liberia}
        \end{subfigure}
        \hspace{-0.3in}
        \begin{subfigure}[b]{0.55\textwidth}
                \includegraphics[width=\textwidth]{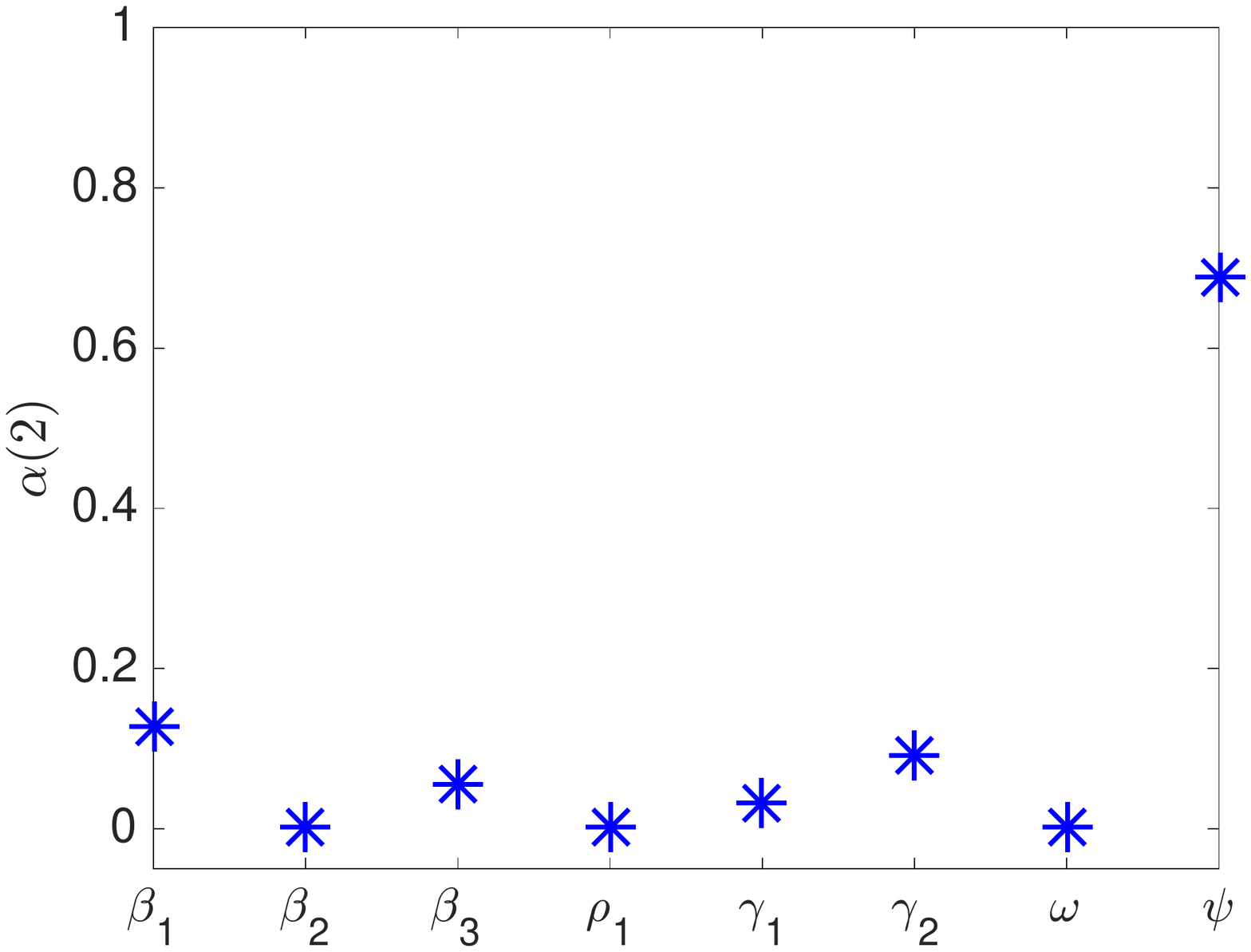} %Liberia_fit_5-23
                \vspace{-1.3in}
                \caption{\footnotesize Sierra Leone}
        \end{subfigure}
        
        \caption{\footnotesize{Activity scores, normalized to $1$. This was computed using $n=2$, the dimension of the active subspace. In both countries, the parameter $\psi$ is by far the most influential on the reproductive ratio. We may also conclude that the parameters $\beta_2,\rho_1$, and $\omega$ have neglible influence on $R_0$ in both Liberia and Sierra Leone. The computation of the matrix $\mC = \mW \mLambda \mW^T$ in \eqref{eq:C} (and therefore the activity scores) is done via tensor product Gauss-Legendre quadrature on $8^8$ points (eight per parameter dimension).} \label{fig:act_scores}}
\end{figure}

\end{document}